\documentclass[draft]{agujournal2019}
\usepackage{url} %this package should fix any errors with URLs in refs.
\usepackage{lineno}
\usepackage{soul}
\usepackage{enumitem}

\usepackage{color}
\draftfalse

\journalname{Journal of Advances in Modeling Earth Systems (JAMES)}

\begin{document}

%% ------------------------------------------------------------------------ %%
%  Title
%
% (A title should be specific, informative, and brief. Use
% abbreviations only if they are defined in the abstract. Titles that
% start with general keywords then specific terms are optimized in
% searches)
%
%% ------------------------------------------------------------------------ %%

% Example: \title{This is a test title}

\title{Physically Interpretable Neural Networks for the Geosciences: Applications to Earth System Variability}

\authors{Benjamin A. Toms\affil{1}\thanks{Department of Atmospheric Science, Colorado State University, Fort Collins, CO}, Elizabeth A. Barnes\affil{1}, Imme Ebert-Uphoff\affil{2,3}}

\affiliation{1}{Department of Atmospheric Science, Colorado State University, Fort Collins, CO}
\affiliation{2}{Department of Electrical and Computer Engineering, Colorado State University, Fort Collins, CO}
\affiliation{3}{Cooperative Institute for Research in the Atmosphere, Colorado State University, Fort Collins, CO}

\correspondingauthor{Benjamin A. Toms}{ben.toms@colostate.edu}

\begin{keypoints}
\item Interpretable neural networks can identify the coherent spatial patterns of known modes of earth system variability
\item The layerwise relevance propagation and backwards optimization (optimal input) methods enable new ways to use neural networks for geoscientific research
\item We propose that the interpretation of what a neural network has learned can be used as the ultimate scientific outcome of a trained network
\end{keypoints}

\begin{abstract}
Neural networks have become increasingly prevalent within the geosciences, although a common limitation of their usage has been a lack of methods to interpret what the networks learn and how they make decisions.  As such, neural networks have often been used within the geosciences to most accurately identify a desired output given a set of inputs, with the interpretation of what the network learns used as a secondary metric to ensure the network is making the right decision for the right reason. Neural network interpretation techniques have become more advanced in recent years, however, and we therefore propose that the ultimate objective of using a neural network can also be the interpretation of what the network has learned rather than the output itself.

We show that the interpretation of neural networks can enable the discovery of scientifically meaningful connections within geoscientific data. In particular, we use two methods for neural network interpretation called backwards optimization and layerwise relevance propagation, both of which project the decision pathways of a network back onto the original input dimensions. To the best of our knowledge, LRP has not yet been applied to geoscientific research, and we believe it has great potential in this area. We show how these interpretation techniques can be used to reliably infer scientifically meaningful information from neural networks by applying them to common climate patterns.  These results suggest that combining interpretable neural networks with novel scientific hypotheses will open the door to many new avenues in neural network-related geoscience research.
\end{abstract}

\section*{Plain Language Summary}

Neural networks, a form of machine learning, have become popular in geoscience over the recent past. A common limitation of neural networks in geoscience has been the belief that they are ``black boxes", and their decision-making process is uninterpretable. This has sometimes made geoscientists hesitant to use neural networks, since an understanding of how and why our models make decisions is important to our science. Methods for interpreting neural networks have become more advanced, however, and so we highlight two such methods that we think have particular promise in geoscientific applications.

The methods are called backwards optimization and layerwise relevance propagation, both of which help identify which inputs into the neural network were most helpful in the neural network's decision-making process. Layerwise relevance propagation has not yet been introduced to the geoscientific community, and we think it offers particularly useful interpretation traits, so we introduce it here. We apply the methods to two commonly studied climate patterns, the El Ni\~no Southern Oscillation and its impacts on seasonal climate patterns over North America, to showcase their utility. Our results suggest that these two interpretation methods open many new avenues for the usage of neural networks within geoscience.

%% ------------------------------------------------------------------------ %%
%
%  TEXT
%
%% ------------------------------------------------------------------------ %%

%%% Suggested section heads:
% \section{Introduction}
%
% The main text should start with an introduction. Except for short
% manuscripts (such as comments and replies), the text should be divided
% into sections, each with its own heading.

% Headings should be sentence fragments and do not begin with a
% lowercase letter or number. Examples of good headings are:

% \section{Materials and Methods}
% Here is text on Materials and Methods.
%
% \subsection{A descriptive heading about methods}
% More about Methods.
%
% \section{Data} (Or section title might be a descriptive heading about data)
%
% \section{Results} (Or section title might be a descriptive heading about the
% results)
%
% \section{Conclusions}

\section{Introduction}

Machine learning methods are emerging as a powerful tool in scientific applications across all areas of geoscience (e.g.~Gil et al., 2018; Kapartne et al., 2018; Rolnick et al., 2019)\nocite{gil2018intelligent,rolnick2019tackling,karpatne2018machine}, including marine science (e.g.~Malde et al., 2019)\nocite{malde2019machine}, solid earth science (e.g.~Bergen et al., 2019)\nocite{bergen2019machine}, and atmospheric science (e.g.~Barnes et al., 2019; Boukabara et al., 2019; Lopatka, 2019; McGovern et al., 2017, Reichstein et al., 2019)\nocite{BoukabaraEos2019,reichstein2019deep,Lopatka2019-dt,mcgovern2017using}.  This revolution in machine learning within the geosciences has been spurred by the coincident introduction of novel algorithms, an influx of large quantities of high-quality data, and an increase in computational power for processing immense quantities of data simultaneously.  There have been limitations to the application of machine learning methods within geoscience, however, as their interpretation is commonly deemed difficult, if not impossible.  Here, we show that two recent techniques from computer science for interpreting one of the most common forms of machine learning methods -- neural networks -- have the potential to transform how geoscientists use machine learning within their research.  More specifically, these methods enable the usage of neural networks for the discovery of physically meaningful relationships within geoscientific data.

Neural networks, also occasionally dubbed ``deep learning" \cite{lecun2015deep}, are one of the most versatile types of machine learning methods and can be used for a broad range of applications within the geosciences.  Such models have been used for time-series prediction (e.g.~Feng et al., 2015; Gardner \& Dorling, 1999)\nocite{gardner1999neural, feng2015artificial}, identifying patterns of weather and climate phenomena within observations and simulations (e.g.~Barnes et al., 2019; Gagne II et al., 2019; Lagerquist et al., 2019, Toms et al., 2019)\nocite{barnes2019, lagerquist2019deep, gagne2019interpretable, toms2019deep}, and parameterizing sub-grid scale physics within numerical models (e.g.~Bolton \& Zanna, 2019; Brenowitz \& Bretherton, 2018; Brenowitz \& Bretherton, 2019; Chevallier et al., 1998; Krasnopolsky et al., 2005; Rasp et al., 2018)\nocite{bolton2019applications, brenowitz2019spatially, rasp2018deep, brenowitz2018prognostic, chevallier1998neural, krasnopolsky2005new}.  The structure of the neural networks employed within these applications can vary substantially, although the general concept is the same: given a set of input variables, the neural network is tasked with identifying the desired output as accurately as possible.  

Neural networks consist of consecutive layers of nonlinear transformations and adjustable weights and biases \cite{goodfellow2016deep}. The mathematics of how these layer-to-layer transformations are applied to the data are well understood since the individual transformations themselves are mathematically simple (e.g.~Sibi et al., 2013)\nocite{sibi2013analysis}.  However, once a neural network has been trained, the reasoning of how and why it combines information across its weights and biases and from each transformation to the next to arrive at its ultimate output is not easily deduced, due to the potentially high complexity of the network architecture and the increasing level of abstraction in later layers of the network \cite{samek2020toward}.  Thus, in practice, neural networks are often used - including in geoscience - without a detailed understanding of the reasoning they employ to arrive at their output.

Even for applications where the network's output is all that is desired, a lack of understanding of a network's reasoning can lead to many problems.  For example, the neural network can overfit to the data and attempt to explain noise rather than capturing the meaningful connections between the input and output. Additionally, within the geosciences sample sizes are typically limited, which means that the available samples might not capture the full range of possible outcomes and thereby might also not be representative of the true underlying physics driving the relationship between the inputs and outputs. In this scenario, the network may fail to model the relationship correctly from a physical perspective, even if it accurately captures a relationship between the inputs and outputs given the provided training data.  Thus, the ability to interpret neural networks is important for ensuring that the reasoning for a network's outputs are consistent with our physical understanding of the earth system.

The various applications of neural networks within the geosciences commonly rely on indirect scientific inference. In many cases, the primary objective of the neural networks has been to maximize the accuracy of the networks' outputs, from which indirect inferences have been made about the earth system. For example, by using neural networks to predict the likelihood that a convective storm would produce hail, Gagne et al. (2019)\nocite{gagne2019interpretable} showed that the neural networks made accurate predictions by identifying known types of storm structures. In another case, Ham et al. (2019)\nocite{ham2019deep} used a neural network to predict the evolution of the El Ni\~no Southern Oscillation (ENSO), and then used interpretation techniques to show that ENSO precursors exist within the South Pacific and Indian Oceans. However, even in these cases, the primary objective was to construct a neural network that most accurately predicted its output, with the interpretation being used to ensure the network attained high accuracy using reasoning consistent with physical theory.  This theme is common throughout geoscientific applications of neural networks: the network's output is the ultimate objective, and interpretation techniques are used to ensure the network is making decisions according to our current understanding of how the earth system evolves. There have also been recent efforts within the geoscience community to compile methods for improving machine-learning model interpretability, including those by McGovern et al. (2019).\nocite{mcgovern2019making} 

We propose an additional use for neural networks, whereby the ultimate scientific objective of using a neural network is its interpretation rather than its output.  From this perspective, we show how neural networks can be used to directly advance our understanding of the earth system.  To do so, we focus on two methods -- backwards optimization and layerwise relevance propagation -- which trace the decision of a neural network back onto the original dimensions of the input image, and thereby permit the understanding of which input variables are most important for the neural network's decisions. These methods are particularly well-suited for scientific inference when a physical understanding of relationships is important, such as within geoscience. We find that layerwise relevance propagation is particularly well suited for geoscientific applications, and has yet to be introduced to the geoscience community to the best of our knowledge.

We first discuss the theory and logic behind the two interpretation methods, then provide two examples of how these methods can be used to explore physically meaningful patterns of earth system variability. The objective of this paper is to showcase the utility of using neural network interpretations for scientific inference. So, we analyze two commonly studied climate phenomena, the El Ni\~no Southern Oscillation and its relationship to seasonal prediction, so that we can first ensure the interpretation methods capture known patterns of geophysical variability before extending into the unknown.

%%%%%%%%%%%%%%%%%%%%%%%%%%%%%%%%%%%%%%
\section{Neural Network Architecture}
\label{sec:neural_network}

%In this section we briefly describe the neural network architecture we use within our examples. However, the interpretation methods we discuss are applicable to other types of neural networks. A more extended review of neural networks and their various forms are available through other resources (e.g.~Gagne et al., 2019; Gers et al., 1999; Simon, 1994)\nocite{gagne2019interpretable, gers1999learning, haykin1994neural}. More extensive details of the neural network architecture are in the appendix.

\begin{sidewaysfigure}
    \centerline{\includegraphics[width=50pc,angle=0]{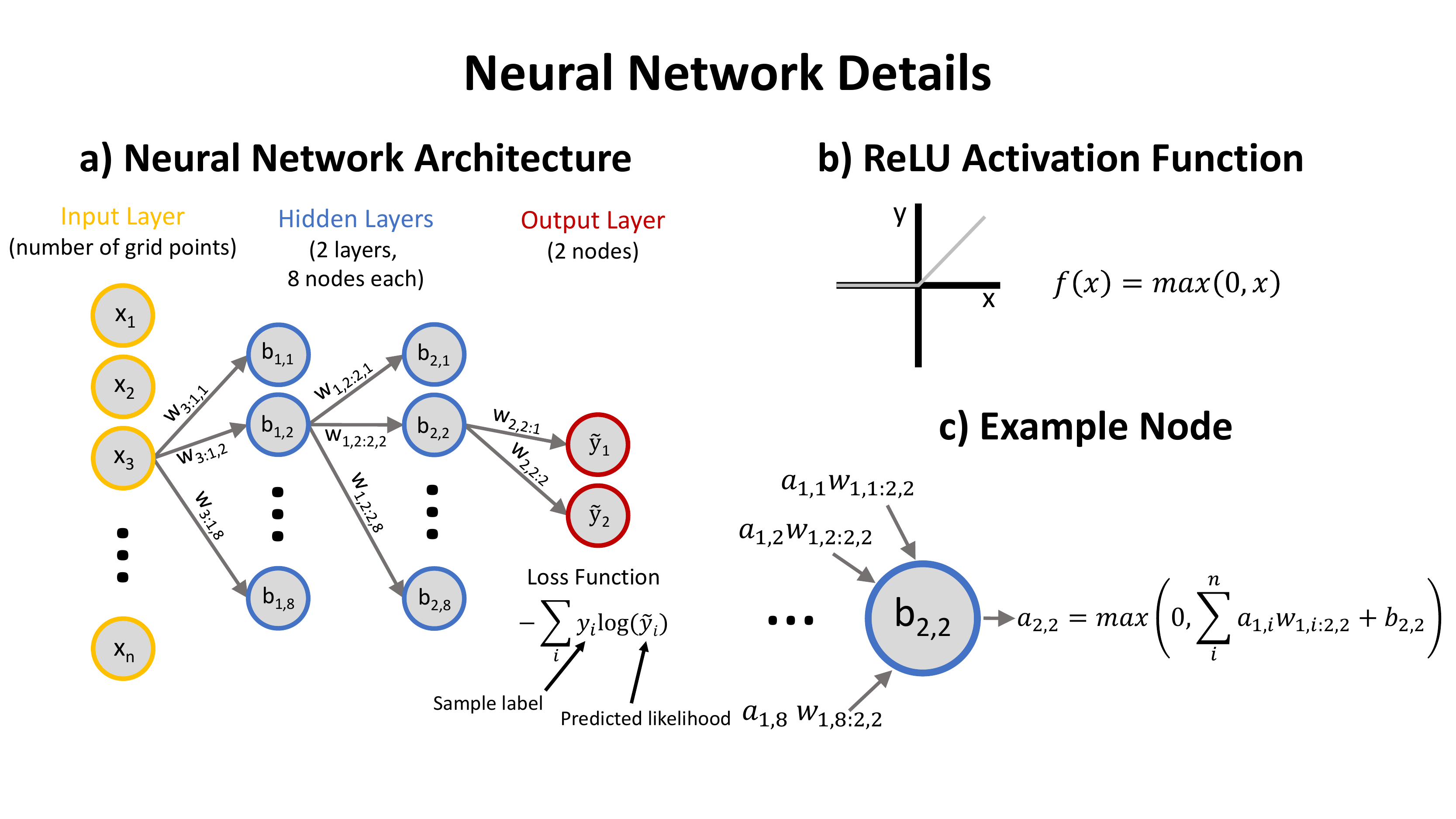}}
    \caption{Illustration of the neural network architecture used in this study.}
    \label{fig:fig1}
 \end{sidewaysfigure}

In this work, we use separately trained fully-connected neural networks of identical design (detailed in Figure \ref{fig:fig1}). A fully-connected neural network is the most basic form of neural network. Each neural network that we use has an input layer which receives the input sample, two intermediate ``hidden" layers of nodes with eight nodes each, and an output layer with two nodes that classifies which of two categories the input is associated with. This type of network is commonly known as a classifier.  The inputs for our examples are vectorized maps (i.e.~images) of geospatial phenomena and are labeled with a two-unit vector that describes which of two categories, or classes, the image is associated with. Within the two-unit labeling vector, a 1 is placed in the index that the sample is associated with and a 0 is placed in the other.  The output of the neural network is also a two-unit vector which represents the neural network's estimation of the likelihood that the input sample belongs in each class such that the output vector always sums to 1, and is calculated using a softmax operator (see appendix for more details). If the neural network is more confident that a sample belongs in a particular class, then the output for the corresponding unit of the output vector will be closer to 1. The objective of the neural network is to output a two-unit vector that is as similar to the label vector as possible, which means it is tasked with maximizing its confidence that each input sample belongs in its labeled category. More extensive details of the neural network architecture and training procedure are provided in the appendix.

It is worth noting that we use a basic form of a neural network for our examples, but could have chosen more advanced architectures such as convolutional neural networks (CNNs, e.g.~Kriszhevsky et al., 2012)\nocite{krizhevsky2012imagenet}. The neural networks we employ are relatively shallow in that they have few layers, whereas it is becoming more common to use ``deep" neural networks with many layers. However, the intent of this paper is to present the usage of the interpretation of neural networks as a tool for scientific inference and not to showcase the utility of various neural network architectures. We therefore opt to keep the networks as simple as possible. In addition, we will show that this basic network architecture is sufficient to capture the known relationships between the inputs and outputs of our examples. The interpretation methods we use also place some restrictions on the structures of the neural networks, the details of which are discussed in the subsequent sections, and so our neural networks abide by these requirements. With that said, the interpretation methods we discuss here are also applicable to a variety of other neural network architectures.

\section{Neural Network Interpretation Methods}
\label{sec:methods}

\subsection{Backwards Optimization (Optimal Input)}
\label{sec:oi}

The technique called backwards optimization calculates the input that maximizes a neural network's confidence in its output, and we therefore refer to the generated pattern as the ``optimal input" \cite{olah2017feature, simonyan2013deep, yosinski2015understanding}. This method offers insights into which patterns the neural network thinks are most associated with a particular output by using the weights and biases of a trained neural network to iteratively update an input sample until it is most closely associated with a user-specified output of the network. 

Once a neural network is trained, the weights and biases can be frozen, which means that they are no longer updated as the neural network sees new samples. So, in turn, the backwards optimization method takes the reverse approach to how a neural network is trained, and rather than updating the weights and biases of the network itself, an input sample is iteratively updated given a trained neural network with frozen weights and biases. The fact that the optimized input has the same dimensions as the samples used to train the network is particularly useful and is helpful for determining which patterns within the input vector are most important for describing any relationships between the input and output variables. The optimized input can also be interpreted in the same units as the input samples used to train the network.

\begin{sidewaysfigure}
    \centerline{\includegraphics[width=50pc,angle=0]{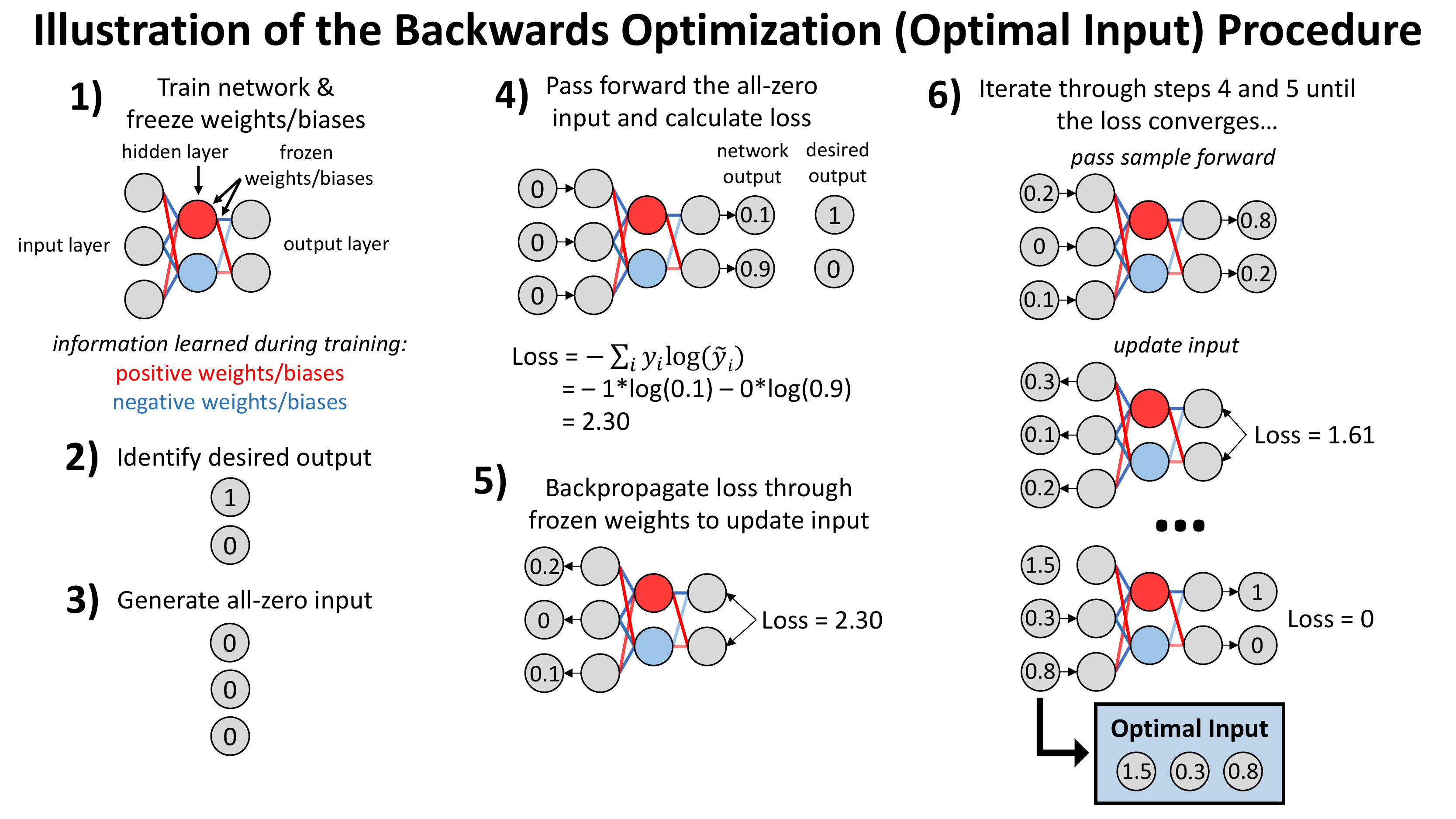}}
    \caption{Illustration of the backwards optimization procedure used in this study for interpreting neural networks. The steps illustrated here correspond to the steps listed in Section \ref{sec:oi}. The neural network within this schematic has already been trained, and the training procedure is not illustrated.}
    \label{fig:fig2}
 \end{sidewaysfigure}

The backwards optimization method is illustrated in Figure \ref{fig:fig2}, detailed in code in the supporting information, and proceeds as follows:
\textit{\begin{enumerate}[align=left]
    \item [Method Input:] User-defined output of a trained neural network
    \item [Method Output:] An optimized input that shows the input pattern most closely associated with the user-defined output according to the trained neural network
    \item [Procedure:]
    \item A neural network is trained, and the weights and biases are frozen, which means that they are not updated when a sample is input into the neural network.
    \item A desired output from the neural network is defined.  For example, if the network is trained to identify whether a sample belongs in one of two categories, the desired output could be when the neural network is 100\% confident that the input belongs in one of the two categories.
    \item A sample is generated of the same shape as the samples used to train the neural network, but the sample is initialized as all zeros.
    \item This all-zero sample is passed through the network, and the output is gathered. The output is then compared to the desired output, and the loss (i.e.~error) of the all-zero sample is calculated with respect to the desired output. The loss function is the same function used to train the network.
    \item The loss is translated backwards through the neural network to the input layer using backpropagation.  But, rather than updating the weights and biases of the network along the way, the input sample itself is updated in a manner which reduces the loss using an increment of the information, or gradient, that was translated back to the input layer.
    \item Iterate over steps 4 and 5 until the input is optimized such that iterations no longer reduce the error of the neural network's output.
\end{enumerate}}

Gagne et al. (2019) and McGovern et al. (2019)\nocite{gagne2019interpretable, mcgovern2019making} provide other examples of how the backwards optimization technique has been used in geoscience, and more specifically meteorology. We note that other techniques for the initialization of the unoptimized input sample have been suggested, such as using Gaussian noise rather than all zeros, but we have found that the optimized patterns are not sensitive to these initialization techniques for our examples.

As will be discussed throughout the remainder of this paper, the backwards optimization technique offers valuable insights into a neural network's decision-making process, but it is not without its limitations. Briefly, the optimized input offers one composite perspective of the patterns the network looks for within the input data. This composite perspective introduces problems when applied to domains where, for example, multiple modes of variability may lead to the same outcome. In these cases, the optimal input may contain a combination of each mode, but will not elucidate how these modes may evolve either independently or in tandem with each other. There are ways that the backwards optimization method can be used for some of these applications too, however, such as by optimizing an actual input sample rather than an all-zero sample toward a target output from the neural network. We do not discuss this application here, but McGovern et al. (2019) briefly discuss such a technique.

Because of the complications of optimizing for a single optimal pattern, it is useful to also understand what information within each input sample is important for the neural network's associated output.  Fortunately, there are methods for interpreting a neural network in this manner, one of which is called layerwise relevance propagation, which we discuss next.

\subsection{Layer-Wise Relevance Propagation (LRP)}
\label{sec:LRP}

While backwards optimization has previously been used by the geoscience community, we are unaware of any published applications of layerwise relevance propagation to geoscientific problems, and so we go into additional detail describing this method. In contrast to the optimal input technique which generates a single optimized input given a desired output, layerwise relevance propagation (LRP) considers one input sample at a time. The form of LRP that we use was introduced to the computer science community by Bach et al. (2015)\nocite{bach2015pixel}. This form of LRP is also referred to as a ``deep Taylor decomposition" of the neural network because of its relationship to Taylor series expansion \cite{montavon2017explaining}, although the more general class of methods is referred to as LRP and we will therefore refer to the method as such. 

For each input sample, LRP identifies the relevance of each input feature for the network's output, and therefore helps isolate which input features are important for a network's output on a sample-by-sample basis. For example, if the input is an image, the resulting output from LRP is a heatmap in the dimensions of the original image that shows the regions of the image which are most important for generating the network's output for that particular sample. It bears repeating that the heatmap is specific to the input sample and so different inputs yield different heatmaps, the patterns of which depend on how the information from that input is transferred through the network as it makes its decision. LRP can be applied to any sample that is of the same dimensions as those used to train the network, even if the neural network did not see the sample during training.

Next, we generally describe how LRP traces the reasoning of a neural network's decision-making process, although we refer the reader to the manuscripts of Bach et al. (2015) and Montavon et al. (2017)\nocite{montavon2017explaining, bach2015pixel} for more details.  We note that while the LRP methods presented by Bach et al.~(2015) and Montavon et al.~(2017)\nocite{montavon2017explaining, bach2015pixel} are one formulation of LRP, new formulations can be developed according to the more general guidelines posed within Bach et al.~(2015)\nocite{bach2015pixel}.

\begin{sidewaysfigure}
    \centerline{\includegraphics[width=50pc,angle=0]{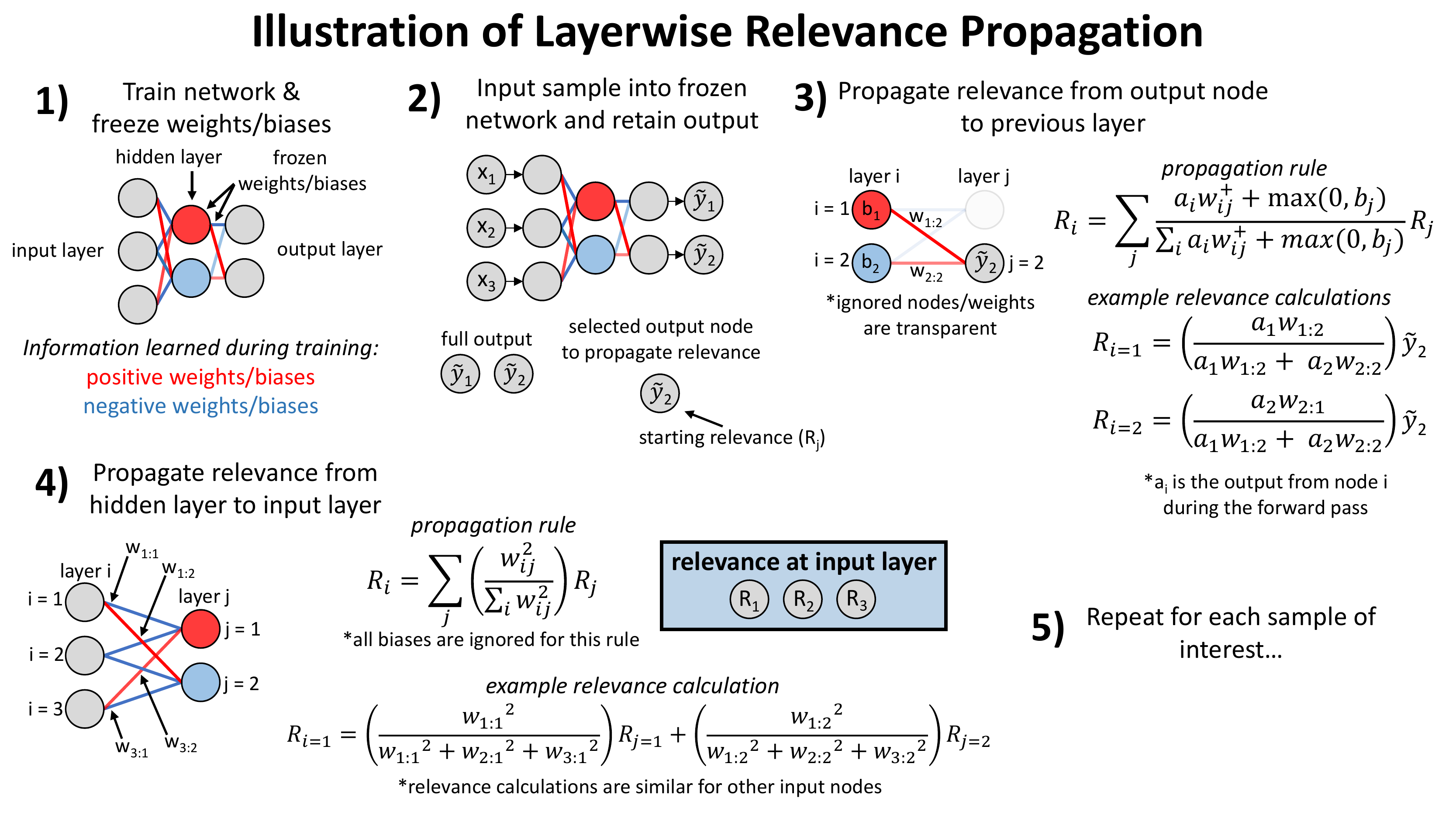}}
    \caption{Illustration of the layerwise relevance propagation (LRP) procedure used in this study for interpreting neural networks. The steps illustrated here correspond to the steps listed in Section \ref{sec:LRP}. The neural network within this schematic has already been trained, and the training procedure is not illustrated. While the illustration does not show the propagation from one hidden layer to another hidden layer, the associated propagation method is identical to the propagation from the output node to the final hidden layer shown in step 3.}
    \label{fig:fig3}
\end{sidewaysfigure}

The algorithm of LRP is illustrated in Figure \ref{fig:fig3} and proceeds as follows:
\textit{\begin{enumerate}[align=left]
    \item [Method Input:] An input sample
    \item [Method Output:] The relevance of each feature within the input sample for the associated output of the neural network
    \item [Procedure:]
    \item A neural network is trained, and the weights and biases are frozen, which means that they are not updated when a sample is input into the neural network.
    \item A sample is then input into the frozen neural network, and the output values are retained. If the neural network has categorical output and uses a softmax operator following the output nodes, then the output values prior to the softmax operator are retained. A single node of the output layer is identified as the node for which the relevance should be calculated. For cases of categorical output, this node is typically the one with the highest output likelihood for the given sample.
    \item The output value of the single node is then propagated backwards through the network using information about the weights and biases of each node of the neural network. The propagation is done according to a particular set of propagation rules, which are discussed below. These rules depend on the types of the neural network and input data, and what type of information is to be inferred from the network.
    \item This backwards propagation through the network is done until reaching the input layer. The resulting values have the same dimensions as the input and correspond to the relevance of each input feature for the neural network's decision of its output.
    \item This process is completed for each sample of interest, from which the relevances for each sample can be studied independently or through composites or clusters of similar patterns of relevance.
\end{enumerate}}

An important aspect of LRP is the rules by which the relevance is translated backwards from the output layer toward the input layer. For our purposes, we only show the relevance propagation rules that are most fundamental to the theory of LRP. The rules that we use here, and which were introduced by Bach et al. (2015)\nocite{bach2015pixel}, have been constructed such that the total summed relevance after propagation back to the input layer is equal to the value of the output. For these rules, only information that $positively$ contributes to the output is propagated backwards, and negative weights and biases are therefore ignored.  That is, only information that makes the network more confident in its categorical output is propagated backwards, and information that makes the network less confident is ignored.  However, there are variants of LRP that permit the inclusion of information that reduces the network's confidence which are also useful for network interpretability, but extend beyond the scope of this paper \cite{montavon2017explaining}.

We note again that LRP traces information for a single output node \cite{bach2015pixel}. So, in the case of categorical output as we present within this paper, the relevance is propagated backwards for one of the categorical output nodes -- typically the node with the maximum output likelihood for the sample of interest. If the neural network uses a softmax operator in its output layer, then during the relevance calculations the softmax operator is ignored and the relevance is calculated for the network's output prior to the softmax. The softmax operator is helpful to ensure the network converges on a solution during training, but the pre-softmax output is more useful for interpretability purposes since it is an unscaled representation of the network's confidence in its output. 

Once a sample has been input, passed forward through the network, and the output has been collected, the first step in LRP is to use the following propagation rule to pass the information backwards from the output layer to the previous layer of nodes:
\begin{equation}
    R_i = \sum_j \frac{a_i w_{ij}^+ + max(0,b_j)}{\sum_i a_i w_{ij}^+ + max(0,b_j)} R_j.
\label{eqn:eqn1}
\end{equation}
Within Equation \ref{eqn:eqn1}, the $i$ subscript represents the $i$-th node in the layer of the network to which the relevance is being translated backwards, the $j$ subscript represents the $j$-th node in the layer of the network from which the relevance is being translated, $R_i$ is the relevance translated backwards to the $i$-th node, $R_j$ is the relevance of the $j$-th node, $a_i$ is the output from the $i$-th node after the non-linearity has been applied when the sample is passed forward through the network, $w_{ij}^+$ is the weight of the connection between the $i$-th and $j$-th nodes where the $+$ signifies that only positive weights are considered, and $b_j$ is the bias of the $j$-th node. The terms within this equation are illustrated schematically within Figure \ref{fig:fig3}. As previously mentioned, the form of LRP that we use neglects all negative weights and biases and only traces information backwards through positive weights and biases. This rule in Equation \ref{eqn:eqn1} is used to propagate the relevance backwards through the network from one layer to the next, starting with the output layer and extending backwards to the first hidden layer.

There are separate rules for translating information to the input layer from the first layer of hidden nodes, the rules of which depend on whether the values of the input features are bounded or unbounded.  A case where the values are unbounded is when the data is standardized and so has zero mean and unit variance, but is not necessarily restricted from varying across all real numbers.  A case where the values are bounded, on the other hand, is when all the input values are normalized between 0 and 1.  For the case where the input values are unbounded, the rule for translating the relevance from the first hidden layer to the input layer is:
\begin{equation}
    R_i = \sum_j \frac{w_{ij}^2}{\sum_i w_{ij}^2}R_j
\label{eqn:eqn2}
\end{equation}
where all terms are as previously discussed for Equation \ref{eqn:eqn1}. 
We use unbounded input data within our examples, and so we provide the propagation rule for the case of bounded data within the supporting information. Additional information about other propagation rules is available within Samek et al. (2019)\nocite{samek2019explainable}.

The rules for LRP presented within the literature have thus far been formulated for a specific subset of activation functions, types of neural networks, and neural network tasks. The rules that we present have been developed to work best with the Rectified Linear Unit (ReLU) activation function, since they test whether a node has been ``activated" or not \cite{bach2015pixel, montavon2017explaining}. Neurons that use the ReLU activation function are activated in the sense that their output is equal to the input if the input is greater than zero, but is zero if the input is less than zero (see Figure \ref{fig:fig1}b for an illustration of the ReLU function). So, the formulation of LRP that we use ensures that it only traces information back through the network if the nodes are activated and therefore pass information forward when the neural network is making its decision for a particular sample. If the $i$-th node is not activated during the forward pass through the network, then the $a_i$ term is zero in Equation \ref{eqn:eqn1}, the relevance for the unactivated neuron $i$ is zero, and the relevance is distributed to the other activated neurons within that layer of nodes.

As we have discussed, we use a form of LRP that only propagates information that positively contributes to the output node, which means that the relevance heatmaps show regions that contribute to increases of the output likelihood that a sample belongs to a particular category. This interpretation is helpful for classification tasks, when increasing the likelihood that an input belongs in a particular category is of interest. There are limitations to this approach for regression problems, however, where it is desirable to understand which inputs cause an increase \textit{or} decrease in the final output. For this reason, we have found that the formulations described by Bach et al. (2015) are not well suited for interpreting neural networks tasked with regression, and we therefore suggest that an LRP formulation needs to be developed specifically for regression problems. However, there have been examples of using LRP for regression problems in other fields (e.g.~Dobrescu et al., 2019)\nocite{dobrescu2019understanding}, and so while LRP may similarly be a viable approach for regression problems in geoscience, care should be taken in how the interpretations are used.

In addition, this formulation of LRP works well for fully-connected neural networks (as we use in this study) and convolutional neural networks, for which the propagation rules are similar \cite{montavon2018methods}. There have been efforts to expand LRP to more complicated neural network architectures, but in these cases other propagation rules need to be used \cite{arras2019explaining}. It is therefore critical that the neural network architecture be carefully considered prior to training if LRP is to be used.

Additional propagation rules for other cases, such as when negative relevances are to be considered, can be found in the supporting information of this paper or within Montavon et al. (2017) and Samek et al. (2019)\nocite{montavon2017explaining, samek2019explainable}.  We use an implementation of LRP from the authors of the method, which is described in detail within Alber et al. (2019)\nocite{innvestigate}, although an abundance of similar implementations also exist.  The implementation we use is available as the $innvestigate$ package within Python, which has been written to work with the $Keras$ neural network package.  Tutorials covering how to implement LRP within other programming languages are available at $heatmapping.org$, and a list of other resources for LRP in $Keras$ and other Python packages is offered within the supporting information.

While there are limitations to LRP, neural networks can be thoughtfully constructed to mediate some of these limitations. For example, many problems of regression can be reformulated as categorical problems by discretizing a continuous output into a number of categories. Additionally, many tasks in geoscience do not seem to require exceedingly complex neural network architectures (e.g.~Barnes et al., 2019; Gagne et al., 2019; Ham et al., 2019)\nocite{barnes2019, gagne2019interpretable, ham2019deep}, and in many cases a basic form of neural network is sufficient to attain high accuracy. Therefore, while the current formulations of LRP do not solve all the limitations of interpreting neural networks for geoscience, we show throughout the remainder of this paper that it still offers opportunities for interpreting neural networks that are thoughtfully constructed with the ultimate objective of interpretation in mind.

%%%%%%%%%%%%%%%%%%%%%%%%%%%%%%%%%%%%
\section{Applications to Earth System Variability}

To illustrate how the interpretation of neural networks can be used to advance scientific knowledge, we apply the backwards optimization and LRP methods to two well-known patterns of climate variability within the earth system.  We intentionally choose patterns that have been extensively researched by the earth system/climate community, because our intent is to demonstrate the usage of neural networks for scientific inference by first showing that the techniques can replicate what we already know before extending into the unknown. Our aim is to provide readers with the intuition and confidence to use the techniques for their own research questions.

For our examples, the inputs to the neural networks are vectorized geospatial fields, the domains of which are discussed in their respective subsections. The neural network is tasked with identifying which of two categories the input geospatial fields are associated with, and what the categories represent depends on the example. It is worth noting that backwards optimization and LRP can be applied to neural networks with any number of output categories, but we limit the output to two categories for the sake of illustration.  Additional details about the neural network architectures we use are discussed in Section \ref{sec:neural_network} and the appendix.

\subsection{The El Ni\~no-Southern Oscillaton (ENSO) Pattern}

%Add in two-three sentences about the time-series of relevance

The first example we use is the simpler of the two, and shows how the backwards optimization and LRP methods can be used to interpret a neural network's understanding of the spatial structure of a well-known climate pattern.  We show that backwards optimization is useful for gaining a composite interpretation of the neural network's understanding of the climate pattern, and that LRP extends beyond this composite and also allows the interpretation of what information is useful to the neural network within each individual sample. This example is intentionally simple so we can test the abilities of the interpretation techniques, rather than gain new knowledge about the climate pattern itself.

\begin{figure}[tp]
    \centerline{\includegraphics[width=25pc,angle=0]{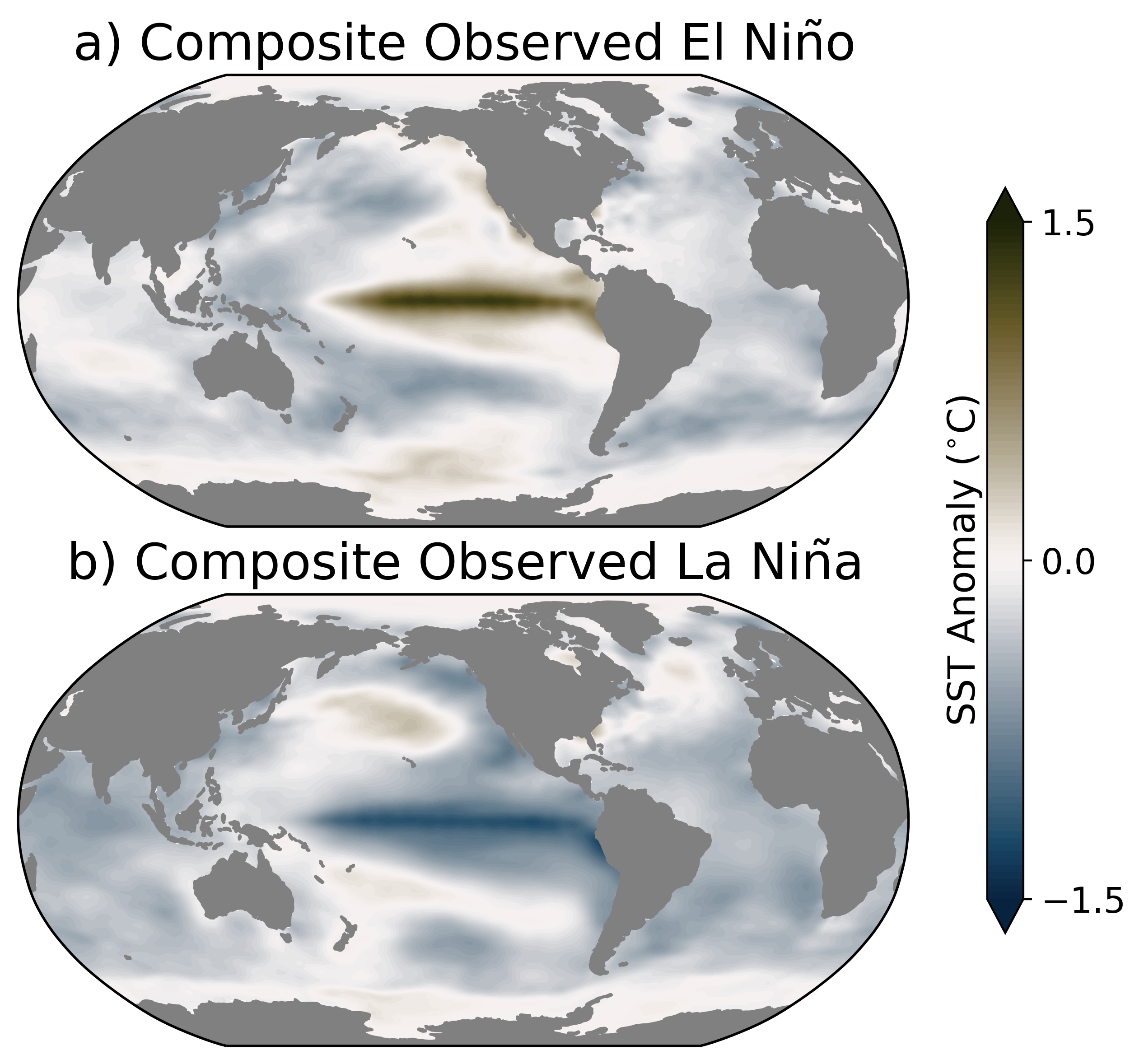}}
    \caption{Composites of the monthly sea-surface temperature anomalies during (a) El Ni\~no (337 samples) and (b) La Ni\~na (485 samples).  The composites include all events with a Ni\~no3.4 index magnitude of greater than 0.5.}
    \label{fig:fig4}
\end{figure}

A neural network is tasked with identifying whether a sea-surface temperature (SST) pattern is characteristic of a positive (El Ni\~no) or negative (La Ni\~na) phase of the El Ni\~no Southern Oscillation (ENSO).  ENSO is a dominant mode of earth-system variability that acts on an interannual timescale and manifests as sea-surface temperature anomalies within the tropical Pacific, although its indirect influences on weather and climate are global \cite{philander1983nino, rasmusson1983meteorological}.  We use the monthly Ni\~no3.4 index downloaded from \textit{https://climatedataguide.ucar.edu/climate-data/} to define the state of ENSO, which is based on an average of sea-surface temperature anomalies within the east-central tropical Pacific (between 5$^{\circ}$S to 5$^{\circ}$N and 170$^{\circ}$W to 120$^{\circ}$W). According to this index, negative sea-surface temperature anomalies within the east-central tropical Pacific are characteristic of La Ni\~na, while positive sea-surface temperature anomalies are characteristic of El Ni\~no. Composite sea-surface temperature anomalies for each phase are shown in Figure \ref{fig:fig4}.

\begin{figure}[tp]
    \centerline{\includegraphics[width=37pc,angle=0]{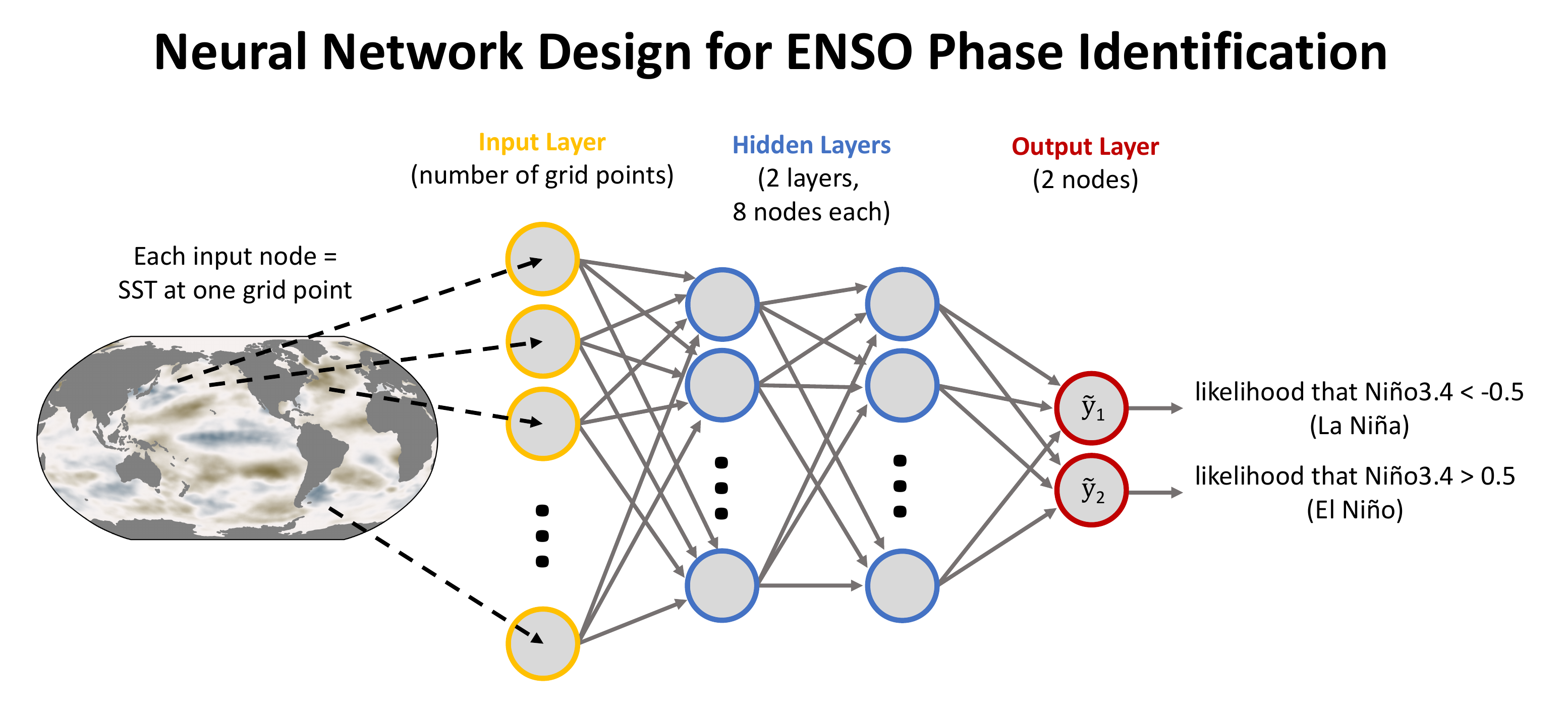}}
    \caption{Illustration of the neural network design for ENSO phase identification.}
    \label{fig:fig5}
\end{figure}

For the neural network setup (shown in Figure \ref{fig:fig5}), the first index of the label vector corresponds to La Ni\~na samples and the second index to El Ni\~no samples. An example vector label for a La Ni\~na case is therefore $[1,0]$, and the output of the neural network is of similar form with the output value in each index corresponding to the network's estimated likelihood that the sample belongs in each category.  The input dataset is monthly sea-surface temperature anomalies for the years 1880 through 2017 from the 1$^{\circ}$ by 1$^{\circ}$ Cobe V2 dataset \cite{hirahara2017centennial}. We calculate the anomalies separately for each grid point by removing the mean for the years 1980 through 2009 and thereafter removing the linear trend. Samples from the years 1880 through 1990 are used to train the network and those from 1990 through 2017 are used to test the network, and we only test and train on months during which the Ni\~no3.4 index magnitude was greater than 0.5. The network does not see the 1990 through 2017 samples during training, and those samples are only used to test whether what the network learns during training generalizes to samples on which the network was not trained. We vectorize the global images of sea-surface temperature anomalies before inputting them into the neural network.

%\begin{sidewaysfigure}[tp]
 %   \centerline{\includegraphics[width=50pc,angle=0]{NinoSchematic.pdf}}
  %  \caption{Illustration of the neural network design for ENSO phase identification.}
 %   \label{fig:fig5}
 %   \end{sidewaysfigure}

We also compare the results to linear regression to verify that the neural network is capturing physically reasonable patterns, since the sea-surface temperature signal of ENSO is predominantly linear although does exhibit nonlinearities \cite{dommenget2013analysis, monahan2001nonlinear}. For this approach, we first obtain a map of regression coefficients by regressing the time series of global sea-surface temperature anomaly maps onto the Ni\~no3.4 index time series.  We then project this map of regression coefficients onto the observed sea-surface temperature anomalies to identify the ENSO phase.

The trained neural network identifies the ENSO phase with 100\% accuracy on both the training (654 samples) and testing (168 samples) datasets. It is expected that the neural network would have nearly perfect accuracy given the intended simplicity of this example, which we use to illustrate the usefulness of the interpretation techniques. Regardless, in order to achieve this accuracy, the weights and biases of the neural network must contain information about the spatial patterns of sea-surface temperature variability characteristic of ENSO. The linear regression approach is accurate for only 81.5\% of samples, which suggests that the nonlinearities within the neural network are important for describing the spatial structure of ENSO (Figure S1).

\begin{figure}[tp]
    \centerline{\includegraphics[width=22pc,angle=0]{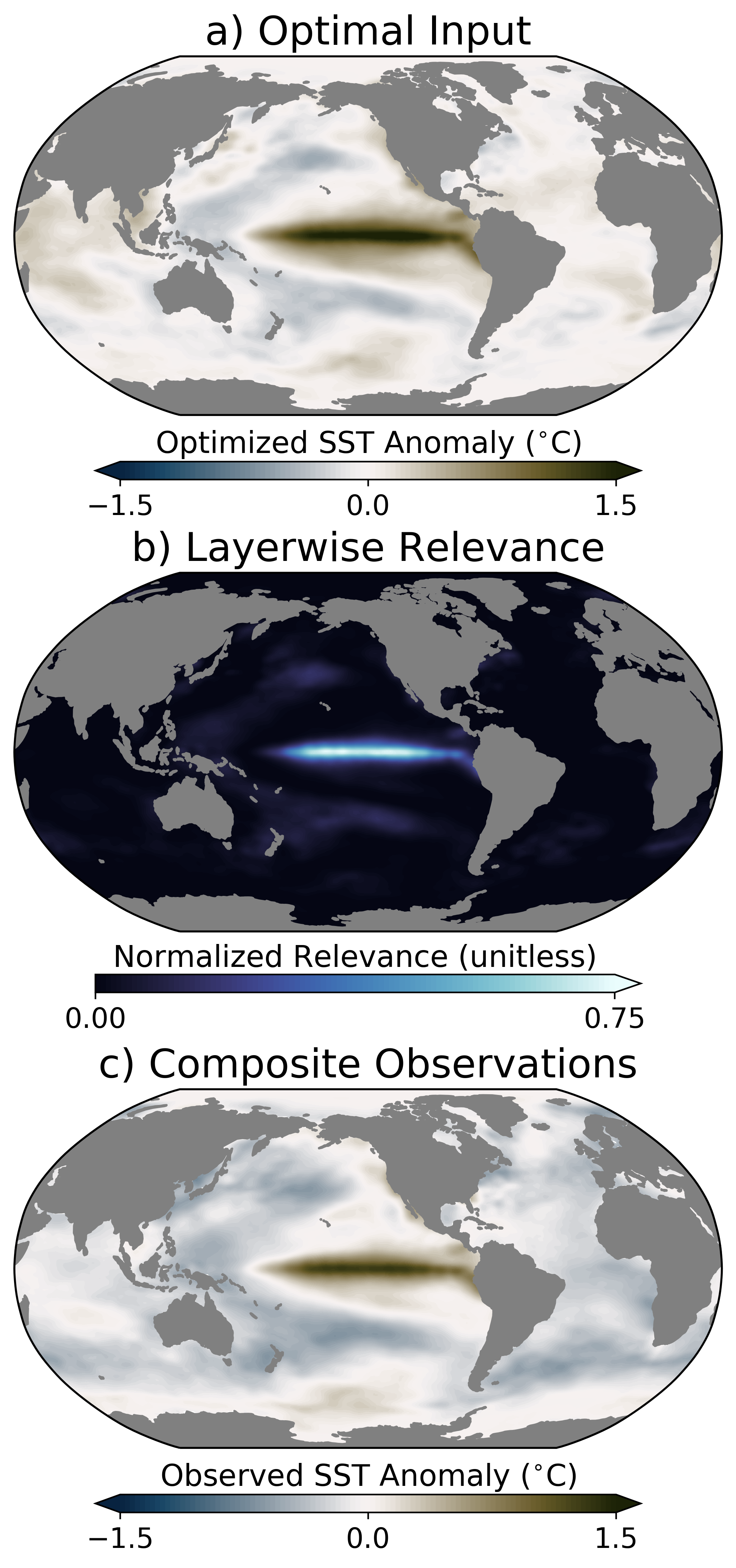}}
    \caption{Interpretation of the neural network's understanding of the spatial structure of El Ni\~no based on 337 total El Ni\~no samples (including both training and testing data).  a) The optimal input field that shows the input image that maximizes the confidence of the network that the sample is an El Ni\~no event.  b) The LRP composite for all El Ni\~no events, where higher values denote greater relevance for the network's decision.  Relevance values are normalized between 0 and 1 for each sample, such that 1 denotes the highest relevance in each individual sample and 0 denotes the lowest relevance. c) Composite observed monthly sea-surface temperature anomalies for all El Ni\~no samples (Ni\~no3.4 $\textgreater$ 0.5), identical to what is shown in Figure \ref{fig:fig4}.}
    \label{fig:fig6}
\end{figure}

\begin{figure}[tp]
    \centerline{\includegraphics[width=33.2pc,angle=0]{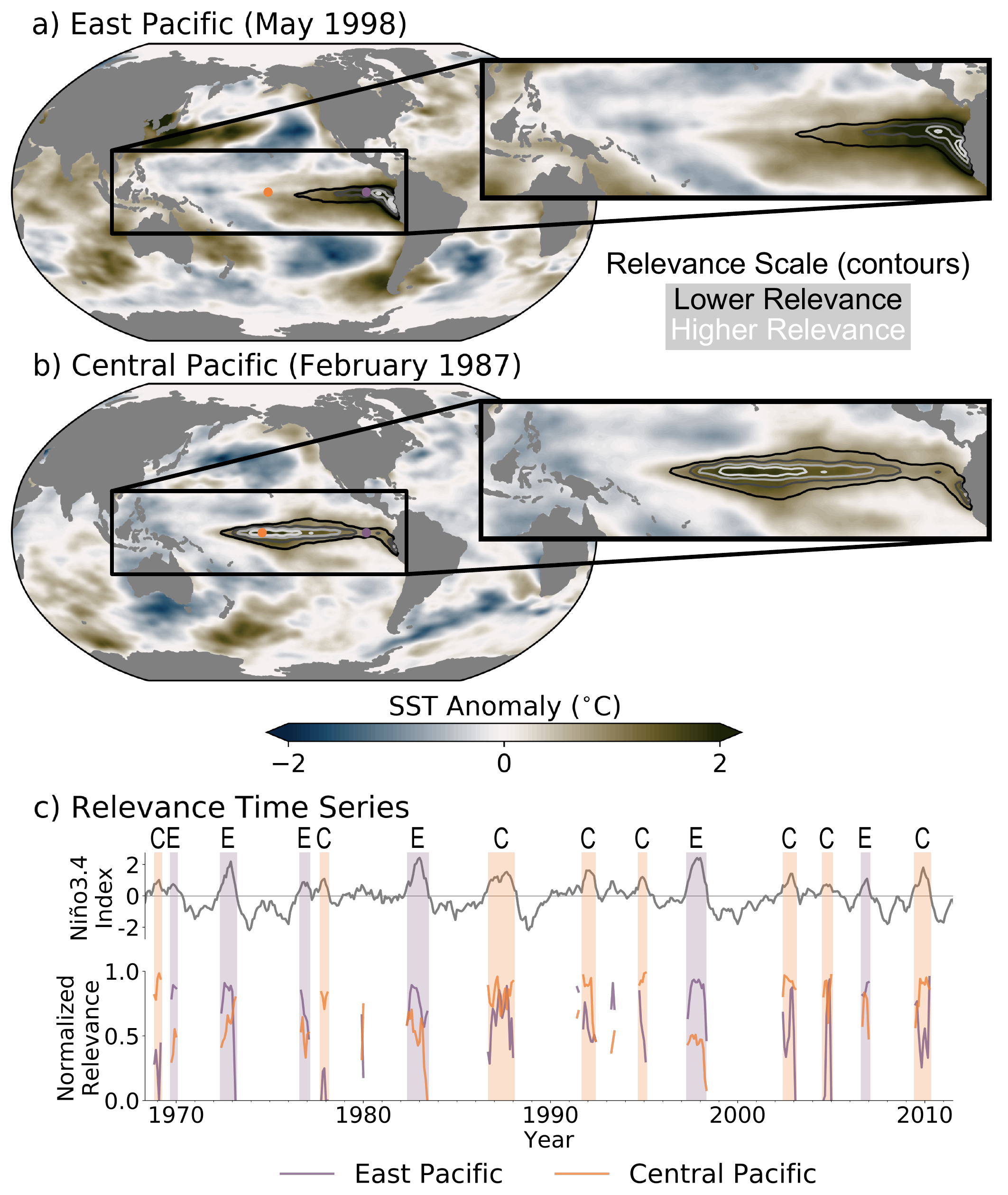}}
    \caption{An illustration of how the neural network focuses on different regions of sea-surface temperature anomalies for different types of El Ni\~no: a) an eastern Pacific El Ni\~no event and b) a central Pacific (Modoki) El Ni\~no event.  The observed sea-surface temperature anomalies for each case are shown in fill and the LRP relevance is contoured. The relevance has been normalized to lie on a scale from 0 to 1, and the contours range in value from 0.2 to 1.0 in increments of 0.2.  Relevance values less than 0.2 have been omitted. c) (top) The Ni\~no3.4 index time series from 1968 to 2011; (bottom) Time series of the normalized relevance values for locations within the central Pacific and eastern Pacific from 1968 through 2011. Relevance values are only shown for months during which the Ni\~no3.4 index was greater than 0.5. The central (eastern) Pacific location is denoted by the orange (purple) dot in panels $a$ and $b$, and is located on the equator at a longitude of 200$^{\circ}$ (250$^{\circ}$). The types of each El Ni\~no event during the 1968 through 2011 period are as labeled in Ashok et al. (2007), Lee and McPhaden (2010), and Wang and Wang (2014)\nocite{ashok2007nino,wang2014different,lee2010increasing}, and are denoted above the time series as either central (``C") or eastern Pacific (``E") events. If an event was not determined to be separable into a central or eastern Pacific event by Ashok et al. (2007), Lee and McPhaden (2010), or Wang and Wang (2014), then it is not labeled.}
    \label{fig:fig7}
\end{figure}

We focus on the interpretation of the neural network's understanding of El Ni\~no, although the interpretation for La Ni\~na is similar and provided in the supporting information (Figure S2). We first generate the optimal input to identify the composite spatial pattern of sea-surface temperature anomalies that maximizes the network's confidence that the sample is an El Ni\~no event (Figure \ref{fig:fig6}a) and the composite relevance heatmaps for all of the El Ni\~no samples (Figure \ref{fig:fig6}b). Then, we use LRP to identify the regions on which the network focuses its attention for El Ni\~no events on a sample-by-sample basis (Figure \ref{fig:fig7}).  The relevance values output from LRP for each sample are normalized to range from 0 to 1 by dividing each heatmap by its own maximum relevance value. We do this so that the relevances for each sample are weighted equally when compositing the relevance across samples.

Backwards optimization recovers a map of sea-surface temperature anomalies that is similar to the observed ENSO pattern in both spatial structure and magnitude, particularly within the tropical Pacific (Figure \ref{fig:fig6}a,c). There are some differences in the sign and magnitude of the anomalies outside of the tropical Pacific, such as in the Atlantic Ocean, although these regions are not conventionally considered to be a part of the predominant ENSO pattern and are also not highlighted to be important to ENSO by the LRP relevance composites (Figure \ref{fig:fig6}b)(e.g.~Philander, 1983)\nocite{philander1983nino}. The composite relevance for the El Ni\~no samples also shows that the neural network mainly focuses its attention on the tropical Pacific (Figure \ref{fig:fig6}b). A region of non-zero relevance exists within the North Pacific (Figure \ref{fig:fig6}b), which may be associated with a well-known correlation between oceanic variability within this region and the tropical signal of ENSO \cite{zhang1996climate}. The linear regression coefficients are spatially similar to the optimal input pattern, which is reassuring given that ENSO is predominantly linear (Figure S1). However, the neural network is more accurate than the regression approach, which implies that the nonlinearities of the neural network are important for identifying the phase of ENSO. The nonlinearities are likely reflected in the uniqueness of each ENSO event, rather than the composite pattern.

%Using a composite of the normalized LRP relevance heatmaps and the optimal input method, we find that the network identifies the canonical signature of El Ni\~no events with remarkable similarity to the observed pattern (Figure \ref{fig:fig4}).  The optimal input field recovers the composite ENSO pattern, and the LRP relevance heatmap shows that the network focuses on the region of sea-surface temperature anomalies in the tropical Pacific.  We also compare the composited sea-surface temperature anomalies for the correctly classified El Ni\~no samples, which is all of the samples, to the derived optimal input (Figure \ref{fig:fig4}c and Figure \ref{fig:fig4}d). The optimal input is similar to the observed El Ni\~no composite, but does exhibit some differences in the sign and magnitude of sea-surface temperature anomalies beyond the tropical Pacific.  However, regions outside of the tropical Pacific are not important for the decision process of the neural network, as shown by LRP, and so it is likely that the lack of attention placed on these other regions reduces the network's ability to identically replicate the full global structure of the sea-surface temperature field.

The utility of LRP is further highlighted by analyzing relevance heatmaps for individual samples.  Figure \ref{fig:fig7}a shows examples of eastern Pacific and central Pacific (i.e.~Modoki; Ashok et al., 2007)\nocite{ashok2007nino} ENSO events in 1998 and 1987 respectively, and highlights that the network refocuses its attention on different regions of the tropical Pacific to identify an El Ni\~no event depending on the input. Furthermore, the neural network focuses its attention on the regions of sea-surface temperature anomalies that are most commonly associated with the two types of El Ni\~no, and learns to ignore other anomalies of similar magnitude within the western Pacific that are distinct from ENSO. We only show the spatial relevance patterns for these two examples, although the relevance time series for the central and eastern Pacific show that the network correctly refocuses its attention for all of the input samples depending on the type of El Ni\~no event (Figure \ref{fig:fig7}c). Samples associated with central Pacific El Ni\~no events have higher relevance within the central Pacific than within the eastern Pacific, and vice versa for samples associated with eastern Pacific El Ni\~no events (Figure \ref{fig:fig7}c).

We have shown that the neural network learns the physical structures of the various modes of ENSO, which lends confidence that backwards optimization and LRP can be used to better our understanding of other patterns of earth system variability. This example also highlights the capability of LRP to identify what information a neural network uses in its decision-making process for each individual sample. The earth system rarely behaves according to a composite, and so the ability to analyze which aspects of each individual sample are important for the neural network's associated output is particularly useful for gaining new insights into earth-system variability.

\subsection{Seasonal Prediction Using the Ocean}

To further illustrate the usefulness of the backwards optimization and LRP methods, we next extend their usage to a slightly more complex example in which we train a neural network to predict a surface temperature response to sea-surface temperature anomalies months in advance. We focus on seasonal prediction, for which the ocean is a predominant source of atmospheric predictability \cite{collins2002climate, doblas2013seasonal, dunstone2011multi}. Specifically, while it is well known that ENSO is a dominant contributor to atmospheric seasonal predictability \cite{ropelewski1986north, wolter1999short}, there are other regions of oceanic variability that offer extended atmospheric predictability. One such region is the North Pacific, which can impact surface temperature and precipitation across North America \cite{capotondi2019predictability, wang2000covariabilities, mckinnon2016long}. We therefore predict continental surface temperature anomalies along the west coast of North America, which is more complicated than predicting the phase of ENSO since the neural network must identify the numerous coincident patterns of sea-surface temperature anomalies across different spatial and temporal scales that can contribute to seasonal temperature predictability.  %We then ask two questions: 1) do the interpretability techniques help identify coherent modes of climate variability that force the temperature response at the location of interest, and 2) .

\begin{figure}[tp]
    \centerline{\includegraphics[width=37pc,angle=0]{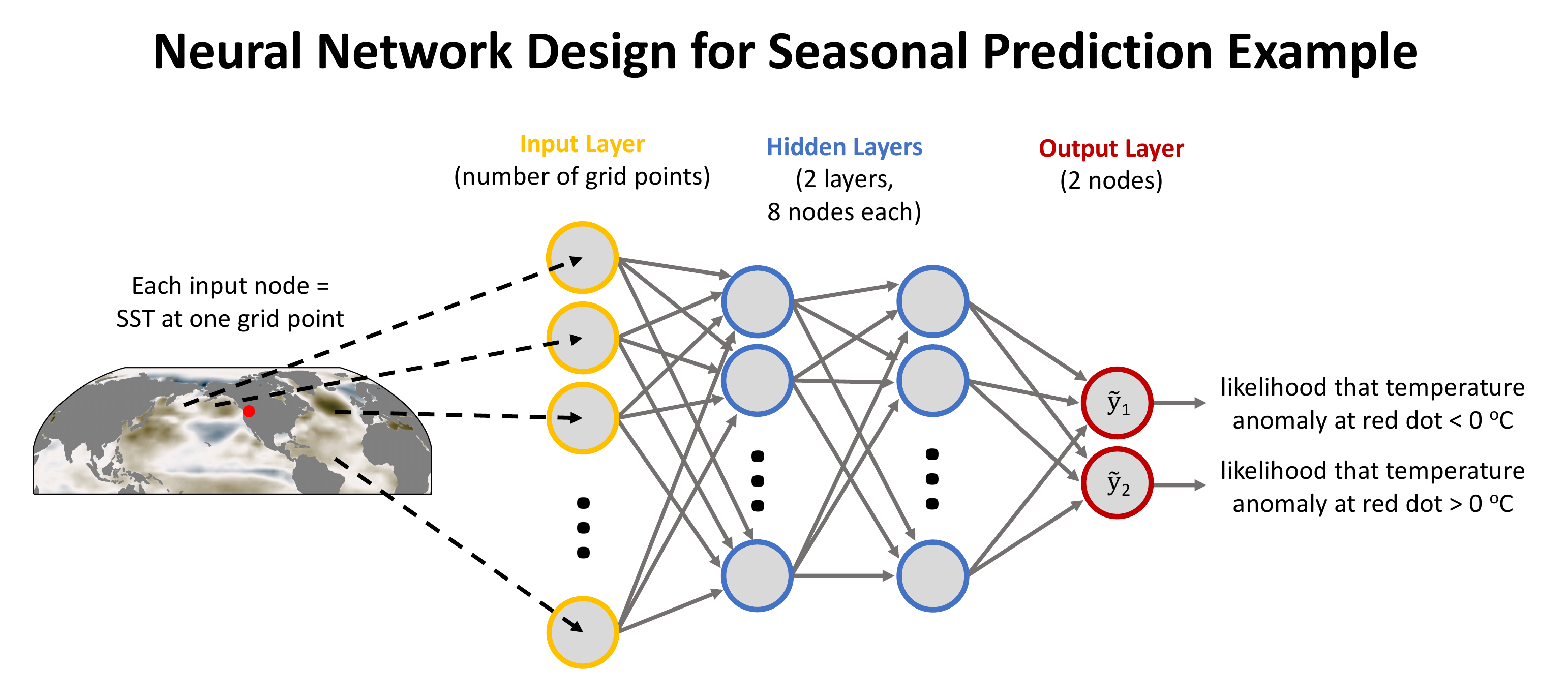}}
    \caption{Illustration of the neural network design for the seasonal prediction example.}
    \label{fig:fig8}
\end{figure}

As shown in Figure \ref{fig:fig8}, we train the neural network to predict the sign (above or below zero) of surface temperature anomalies at a location along the west coast of North America (50$^{\circ}$N, 240$^{\circ}$E) using maps of sea-surface temperature anomalies within the tropics and Northern Hemisphere (north of 20$^{\circ}$S). Surface temperatures at the chosen location, which is denoted by the red dot in subsequent figures, have previously been shown to have extended predictability due to sea-surface temperature forcing on seasonal to annual timescales (e.g.~Capotondi et al., 2019; Gershunov, 1998)\nocite{capotondi2019predictability, gershunov1998enso}. We input sea-surface temperature anomalies from the 1$^{\circ}$ by 1$^{\circ}$ Cobe V2 monthly sea-surface temperature anomaly dataset that is linearly interpolated onto a daily basis \cite{hirahara2017centennial}, and we use the years 1950 to present day. The corresponding daily surface temperature anomaly labels are gathered from the Berkeley Earth Surface Temperatures (BEST; Rohde et al., 2013) dataset, also spanning from 1950 to present day\nocite{rohde2013new}. For both the sea-surface and continental surface temperatures, we calculate the anomalies separately for each grid point by subtracting the mean values for the years 1980 through 2009 and thereafter removing the linear trend.  The training dataset spans from 1950 through 2000 ($\sim$18,000 samples), and the testing dataset spans from 2000 through 2018 ($\sim$7,000 samples). The surface temperature anomalies are averaged over a 60-day period to ensure the predictions are capturing longer-term surface temperature variability, and the averages are centered such that a prediction with a lead time of 60 days implies a prediction of the average 30- to 90-day surface temperature anomalies.

\begin{figure}[tp]
    \centerline{\includegraphics[width=26pc,angle=0]{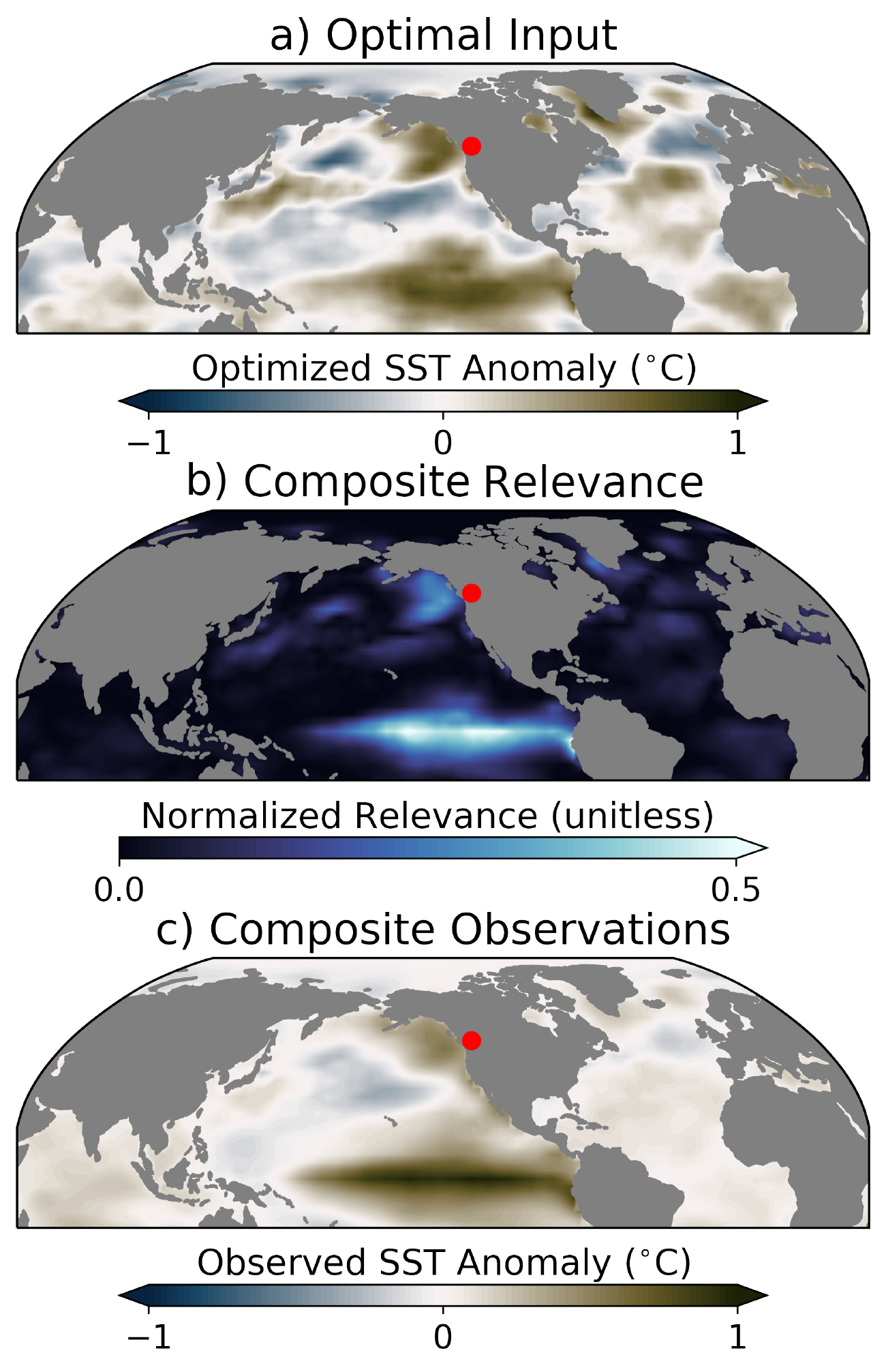}}
    \caption{Interpretation of the neural network tasked with predicting 30- to 90-day average surface temperature anomalies at the red dot based on $\sim$12,000 total samples (including both training and testing data). Only the interpretation for positive surface temperature anomalies is shown, and the interpretation for negative anomalies is shown in Figure S3. a) The optimal input field that maximizes the network's confidence that the input sample is associated with positive temperature anomalies at the red dot.  b) The LRP composite for all correctly categorized samples of positive temperature anomalies, where higher values denote greater relevance.  Relevance values are normalized between 0 and 1 for each sample, such that 1 denotes the highest relevance in each individual sample and 0 denotes the lowest relevance. c) Composite observed sea-surface temperature anomalies for all cases where the neural network accurately predicts positive surface temperature anomalies.}
    \label{fig:fig9}
\end{figure}

We use interpretations of the neural network to identify which sea-surface temperature patterns are useful for making extended surface temperature predictions at various prediction lead times. We first train a neural network to predict the sign of the 30- to 90-day average surface temperature anomalies (i.e. a 60-day lead time using our definition), for which the network has 67\% accuracy. We then focus on interpreting the neural network for cases when the surface temperature anomalies are positive, although the interpretation for the cases with negative anomalies is similar and provided within the supporting information (Figure S3). For this lead time, the optimal input and LRP composite identify similar regions of SST patterns that lend predictability across the tropical Pacific and North Pacific (Figure \ref{fig:fig9}a,b).  Both of these regions have been identified by previous studies as sources of seasonal temperature predictability for the west coast of North America \cite{capotondi2019predictability, gershunov1998enso, wolter1999short}.

\begin{figure}[tp]
    \centerline{\includegraphics[width=35pc,angle=0]{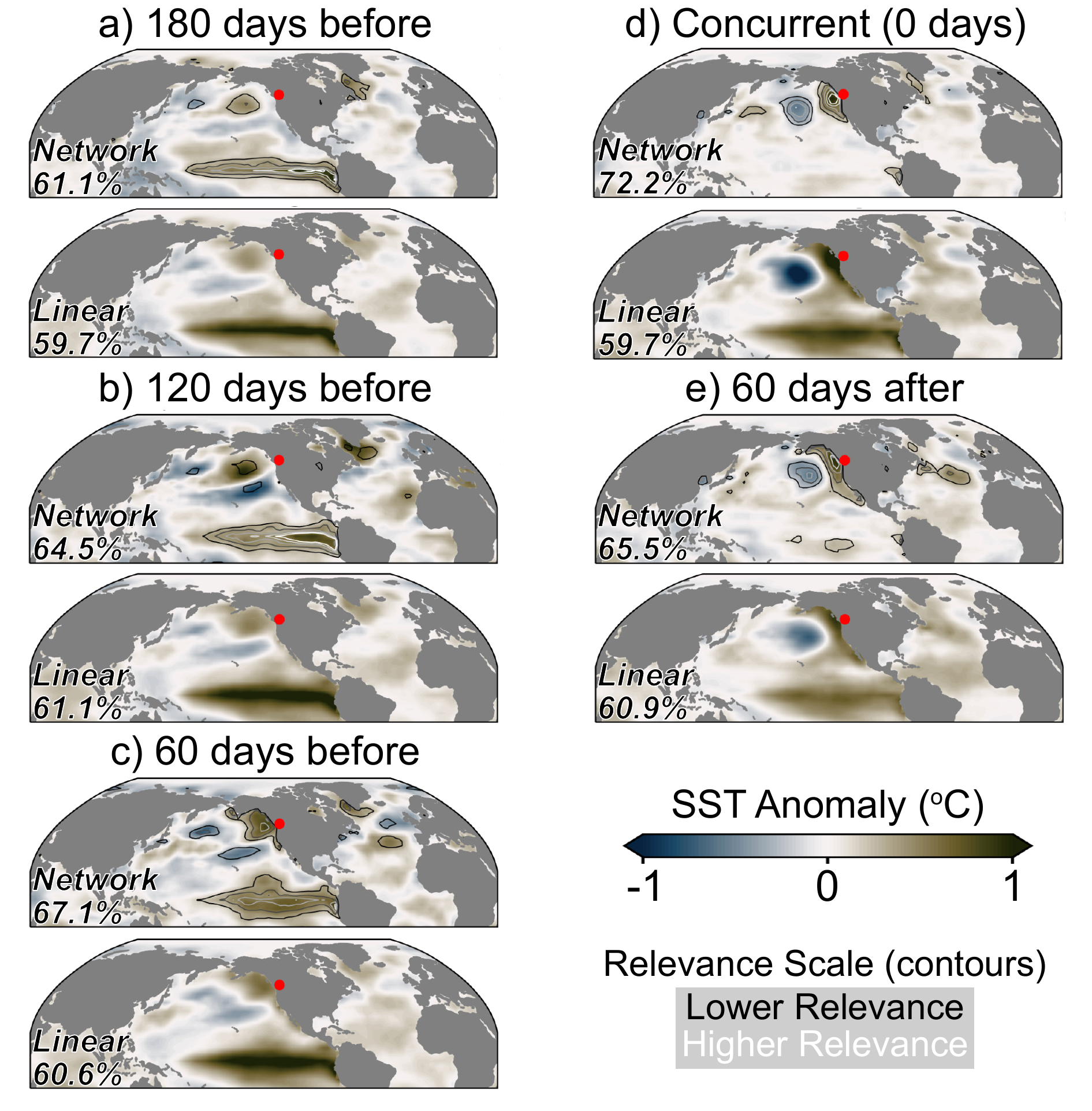}}
    \caption{A comparison of the spatial patterns of sea-surface temperature deemed important for predicting surface temperature at the red dot using neural networks and linear regression.  An evolution of the sea-surface temperature patterns at various lead times is shown, ranging from 180 days prior to the surface temperature anomalies to 60 days afterwards. The prediction is made for surface temperatures averaged across a 60-day window, and the prediction lead time listed above the sub-figures is the center of this window. So, for example, the 180-day lead time prediction is actually a prediction of the 150- to 210-day average surface temperature. For each lead/lag, the top panel shows the neural network optimal input in fill and LRP relevance in open contours, and the bottom panel shows the regression coefficients for the linear regression approach.  The open contours denote LRP relevance values ranging from 0.1 to 0.3 in increments of 0.05.}
    \label{fig:fig10}
\end{figure}

We next test the fidelity of the neural network interpretations by varying the prediction lead time of the continental surface temperature anomalies from 180 days prior to 60 days following their occurrence.  We compare the neural network interpretations with that of linear regression to test whether the interpretations are reliable and if they offer any unique insight compared to more conventional approaches.  Our linear regression approach is similar to the approach used for the ENSO example.  We first obtain a map of regression coefficients by regressing the time series of global sea-surface temperature anomaly maps onto the time series of surface temperature anomalies over the west coast of North America.  We then project the regression coefficient map onto the global maps sea-surface temperature anomalies to predict the sign of the surface temperature anomaly. The resulting accuracies of both prediction methods and the associated sea-surface temperature patterns that lend predictability are shown in Figure \ref{fig:fig10}.

At extended leads, the spatial patterns of sea-surface temperature anomalies identified by backwards optimization and LRP are similar to those identified by regression (Figure 10).  Particularly, the tropical Pacific stands out as being a predominant source of surface temperature predictability across the 180-day, 120-day, and 60-day prediction lead times for both the neural network interpretation and the regression maps (Figure \ref{fig:fig10}a,b,c).  For the 60-day prediction lead time, within the neural network interpretations the importance of the North Pacific begins to increase relative to the ENSO region, and the North Pacific becomes the dominant source of predictability for the concurrent and 60-day lagged sea-surface temperature anomalies (Figure \ref{fig:fig10}c,d,e). Unlike the neural network, the regression approach continues to highlight the tropical Pacific Ocean as important for identifying the concurrent and 60-day lagged surface temperature anomalies. 

The neural network is more accurate than the regression approach for all prediction ranges, which suggests that the neural network interpretations likely capture the sea-surface temperature patterns more closely associated with the seasonal surface temperature anomalies. Specifically, the neural network interpretations suggest that the North Pacific is the predominant modulator of concurrent surface temperature anomalies along the west coast of North America, while the tropical Pacific offers extended lead predictability (Figure \ref{fig:fig10}). This idea is corroborated by previous research that found the North Pacific modulates temperatures across western North America separately from the tropical Pacific \cite{capotondi2019predictability}. 
So, while the neural network is only slightly more accurate than the linear regression model, the increase in accuracy is caused by an improved understanding of the most relevant sea-surface temperature patterns. Either nonlinearities or the increased pathways for information to flow through the neural network likely contribute to this improved understanding.

%This further suggests that the North Pacific is the predominant modulator of concurrent surface temperature anomalies along the west coast of North America, but that the tropical Pacific offers extended lead predictability. It is plausible that tropical Pacific sea-surface temperature anomalies preclude sea-surface temperature anomalies within the North Pacific, although we leave a more thorough discussion of this for another paper. The North Pacific has been identified as a local forcing mechanism for temperature anomalies along the west coast of North America by previous studies, and while tropical Pacific anomalies do commonly occur during such instances of extended predictability, oftentimes North Pacific sea surface temperature anomalies are ultimately responsible for the actual surface temperature forcing \cite{capotondi2019predictability}. The interpretation of the neural network is able to capture this concept, and would encourage the scientist to look beyond the tropical Pacific in their mechanistic explanations for surface temperature predictability.

\section{Discussion and Conclusions}

The recent surge in the popularity of neural networks within the geosciences has inspired the need for techniques to interpret their decisions.  Neural networks are conventionally thought of as ``black boxes" within the geosciences with limited tools for the interpretation of the reasoning behind their decision-making process. We have shown that the usage of two separate techniques enables physically meaningful inference from thoughtfully designed neural networks.  This ability to reliably interpret neural networks opens the door to using the interpretation of how and why the network makes its decisions as the ultimate science outcome.

The backwards optimization method can be used to quantify the patterns within the input data that maximize a neural network's confidence that an input is associated with a particular output.  For the case of categorical output as we present within this paper, backwards optimization iteratively changes an input to maximize the neural network's confidence that it belongs in a particular category. The optimized input has the same dimensions and can be interpreted in the same units as the input samples used to train the network, but provides no direct indication as to which characteristics of the optimized input are most important. In general, however, backwards optimization is useful for identifying the dominant pattern of variability the neural network looks for when making its decisions. In our examples of ENSO phase identification and seasonal prediction, backwards optimization was able to extract the dominant modes of variability known to be associated with each problem (Figure \ref{fig:fig6}; Figure \ref{fig:fig10}).

Layerwise relevance propagation (LRP), on the other hand, considers each sample individually, and provides information about the characteristics of each sample that are most important, or relevant, for the network's associated output.  LRP can thereby provide insights into how relationships between the inputs and outputs of a neural network vary on a case-by-case basis. The usefulness of this quality is exemplified by comparing the relevance heatmaps for two types of El Ni\~no events -- the eastern Pacific and central Pacific, or Modoki, patterns (Figure \ref{fig:fig7}).  Although the optimal input pattern does not distinguish between these two modes of El Ni\~no variability because it offers a composite interpretation (Figure \ref{fig:fig6}a), LRP shows that the network does redirect its focus depending on where the sea-surface temperature anomalies occur (Figure \ref{fig:fig7}). While we do not examine this capability within this paper, it is possible to cluster the LRP relevance heatmaps to identify secondary modes variability within each input category if there is no a-priori knowledge of their existence \cite{lapuschkin2019unmasking}. The fact that the neural network learns the variable spatial structures of ENSO, and that LRP can elucidate this understanding, suggests that LRP can be used to identify physically meaningful patterns within other geoscientific datasets, as well.

%The seasonal prediction problem highlights the ability of interpretable neural networks to advance our understanding of more complex geophysical patterns. The increased complexity of the neural network relative to a linear regression model enables the neural network to identify the most relevant sea-surface temperature patterns for seasonal temperature anomalies more consistently with physical theory (Figure \ref{fig:fig10}). So, even for predominantly linear problems such as ENSO phase identification and the seasonal teleconnections of ENSO, interpretations of neural networks can offer more reliable physical insights than conventional linear methods. Additionally, the usage of LRP enables an evaluation of the patterns most relevant to predictions on a sample-by-sample basis, which is particularly useful for identifying sub-modes of variability , as was shown in the ENSO example. And, finally, while we do not examine this capability within this paper, it is possible to cluster the LRP relevance heatmaps to identify modes of variability within the input samples \citep{lapuschkin2019unmasking}.

There are particular requirements of the backwards optimization and LRP techniques that constrain how a neural network is constructed, the details of which are discussed in Section \ref{sec:methods}. We therefore emphasize that neural networks must be constructed thoughtfully so as to maximize the scientific value of their interpretation. The network architecture must be complex enough to capture any existing relationships between the input and output data, but not so complex that interpretation methods are no longer usable, the balance of which depends on the use-case. The relative value of the accuracy and interpretability of a neural network is of critical importance to scientific analyses, and should be assessed carefully prior to training. For example, first training a simple neural network and building towards a more complex model enables an understanding of whether more complex and thereby less interpretable networks are necessary. If a network is too simple to accurately capture the relationships between the input and output, then its accuracy will be low and any interpretations of its understanding will be limited in scientific value. On the other hand, if a network is too complex and interpretation is impossible, then its value is limited solely to its output. A balance between network complexity and interpretability must be struck if the interpretation of what a network has learned is to be scientifically useful.

We have shown that techniques for interpreting neural networks have the potential to extend their usage to the discovery of unknown patterns within geoscientific data, a concept which will be further explored in future research. The ultimate scientific outcome of a neural network can now also be the interpretation of what the neural network has learned, rather than only the output of the network itself. Regardless of the specific application, it is now apparent that neural networks offer scientists a useful new way to discover and understand connections within geoscientific data.

%Appendix: Add in L2 regularization information

% The \appendix command resets counters and redefines section heads
%
% After typing \appendix
%
%\section{Here Is Appendix Title}
% will show
% A: Here Is Appendix Title
%
%\appendix
%\section{Here is a sample appendix}

\appendix

\section{Additional Neural Network Details}

The individual grid cells within the vectorized inputs, which are maps in our cases, are each treated as independent inputs of the neural network. Each input node receives the value for one element of the input vector and is connected to each node within the first hidden layer of internal nodes.  The individual nodes of the first hidden layer are therefore each connected to every element of the input vector, and can use information from any input element according to the weight connecting the node to the inputs.  The first hidden layer is then connected to the second hidden layer in a similar fashion, with each node within the first hidden layer connected to each node within the second hidden layer. The Rectified Linear Unit (ReLU) activation is applied to the output from each of the hidden layer nodes before the output is passed on to the next layer. Each node within the second hidden layer is finally connected to the two output nodes, which represent the neural network's estimated likelihood that the input sample corresponds to each of the two categories. The weights and biases are initialized randomly using the ``He normal" technique \cite{he2015delving}, such that the they do not contain any information about the relationship between the inputs and outputs upon initialization. When the neural network is trained, the weights and biases of the network are iteratively updated until the output of the network is most similar to the input labels (i.e.~the model is most accurate) once the network's weights and biases have converged on an optimal solution.

The likelihood output is generated by applying a ``softmax" operator to the output of the neural network before estimating its accuracy, which is formulated as follows:
\begin{equation}
    \tilde{y}_i = \frac{exp(x_i)}{\sum_j exp(x_j)}
\end{equation}
where $x_i$ represents the pre-softmax output of the neural network for output node $i$ (of which there are two in our architecture), the numerator is the exponential of the value of that output node, and the denominator is the sum of the exponential of all pre-softmax outputs.  In this sense, the post-softmax output of the neural network is a relative likelihood that the input sample belongs to each class, with higher values being indicative of a higher likelihood, and vice versa.  Following the application of the softmax operator, we then use the cross-entropy loss function to estimate the accuracy of the network, which takes the form of:
\begin{equation}
    loss = - \sum_i y_i log(\tilde{y}_i)
\end{equation}
where $i$ represents the $i$-th unit of the label vector for the input sample, $y_i$ is the value of the $i$-th unit of the label vector, and $\tilde{y}_i$ is the output value of the $i$-th node of the output layer from the neural network after being transformed by the softmax operator.  This loss function therefore assigns error to the output of the neural network on a logarithmic scale based on how different the output likelihood vector is from the label of the input sample, and punishes large errors more severely than small errors due to the logarithmic transformation.

The neural networks are trained using gradient descent with the Nesterov accelerated stochastic gradient descent optimizer \cite{nesterov1983method, ruder2016overview}.  The learning rate is set to an initial value of 0.01 with a Nesterov momentum parameter of 0.9.  The learning rate is reduced by a factor of 0.5 after 50 epochs, and the neural networks are trained for a total of 100 epochs, which is sufficient for convergence for both examples within this paper.

We use L$_2$ (i.e.~ridge) regularization for each example to ensure the network divides its attention across a greater number of input nodes than it otherwise would. For the ENSO problem, we use an L$_2$ parameter of 25 for the weights between the input layer and the first hidden layer and 0.01 for all other weights. For the seasonal prediction problem, we use an L$_2$ parameter of 10 for the weights between the input layer and the first hidden layer and 0.01 for all other weights. We find that a careful selection of the L$_2$ parameter is important for ensuring that the neural network does not overfit to the input data, although our conclusions are consistent for L$_2$ parameters of 5 to 50 between the input layer and first hidden layer.

A more extended review of neural networks and their various forms are available through other resources (e.g.~Gagne et al., 2019; Gers et al., 1999; Goodfellow et al., 2016; Simon, 1994)\nocite{gagne2019interpretable, gers1999learning, haykin1994neural, goodfellow2016deep}.

\acknowledgments
Benjamin A. Toms was funded by the Department of Energy Computational Science Graduate Fellowship via grant DE-FG02-97ER25308. Elizabeth A. Barnes was funded, in part, by NSF-AGS CAREER grant AGS-1749261. Support was also provided to Imme Ebert-Uphoff and Elizabeth A. Barnes through NSF grant 1934668 of the HDR (Harnessing the Data Revolution) program. The Cobe V2 sea-surface temperature data used in this study can be accessed via NOAA ESRL (https://www.esrl.noaa.gov/psd/data/gridded/data.cobe2.html), and the ENSO Ni\~no3.4 index data can be accessed via the NCAR climate data guide \\\noindent(https://climatedataguide.ucar.edu/climate-data/).

%% ------------------------------------------------------------------------ %%
%% References and Citations

%%%%%%%%%%%%%%%%%%%%%%%%%%%%%%%%%%%%%%%%%%%%%%%
%
% \bibliography{<name of your .bib file>} don't specify the file extension
%
% don't specify bibliographystyle
%%%%%%%%%%%%%%%%%%%%%%%%%%%%%%%%%%%%%%%%%%%%%%%

\clearpage
\bibliography{TomsEtAl_2019_InterpretableForGeoscience}

%Reference citation instructions and examples:
%
% Please use ONLY \cite and \citeA for reference citations.
% \cite for parenthetical references
% ...as shown in recent studies (Simpson et al., 2019)
% \citeA for in-text citations
% ...Simpson et al. (2019) have shown...
%
%
%...as shown by \citeA{jskilby}.
%...as shown by \citeA{lewin76}, \citeA{carson86}, \citeA{bartoldy02}, and \citeA{rinaldi03}.
%...has been shown \cite{jskilbye}.
%...has been shown \cite{lewin76,carson86,bartoldy02,rinaldi03}.
%... \cite <i.e.>[]{lewin76,carson86,bartoldy02,rinaldi03}.
%...has been shown by \cite <e.g.,>[and others]{lewin76}.
%
% apacite uses < > for prenotes and [ ] for postnotes
% DO NOT use other cite commands (e.g., \citet, \citep, \citeyear, \nocite, \citealp, etc.).
%

\end{document}

%% ------------------------------------------------------------------------ %%
%
%  TITLE
%
%% ------------------------------------------------------------------------ %%

%\includegraphics{agu_pubart-white_reduced.eps}

\title{Supporting Information for ``Physically Interpretable Neural Networks for the Geosciences: Applications to Earth System Variability"}
%
% e.g., \title{Supporting Information for "Terrestrial ring current:
% Origin, formation, and decay $\alpha\beta\Gamma\Delta$"}
%
%DOI: 10.1002/%insert paper number here%

%% ------------------------------------------------------------------------ %%
%
%  AUTHORS AND AFFILIATIONS
%
%% ------------------------------------------------------------------------ %%

% List authors by first name or initial followed by last name and
% separated by commas. Use \affil{} to number affiliations, and
% \thanks{} for author notes.
% Additional author notes should be indicated with \thanks{} (for
% example, for current addresses).

% Example: \authors{A. B. Author\affil{1}\thanks{Current address, Antartica}, B. C. Author\affil{2,3}, and D. E.
% Author\affil{3,4}\thanks{Also funded by Monsanto.}}

\authors{Benjamin A. Toms\affil{1}\thanks{Department of Atmospheric Science, Colorado State University, Fort Collins, CO}, Elizabeth A. Barnes\affil{1}, Imme Ebert-Uphoff\affil{2,3}}

\affiliation{1}{Department of Atmospheric Science, Colorado State University, Fort Collins, CO}
\affiliation{2}{Department of Electrical and Computer Engineering, Colorado State University, Fort Collins, CO}
\affiliation{3}{Cooperative Institute for Research in the Atmosphere, Colorado State University, Fort Collins, CO}

\section*{}

\noindent\textbf{Contents of this file}
%%%Remove or add items as needed%%%
\begin{enumerate}
\item Text S1 and S2
\item Code S1
\item Figures S1 and S2
\end{enumerate}
\noindent\textbf{Additional Supporting Information (Files uploaded separately)}
\begin{enumerate}
\item Python scripts for the ``Optimal Input" method
\end{enumerate}

\noindent\textbf{Introduction}

Within this supplemental material, we list additional rules for propagation, resources for implementing and using LRP in neural network packages other than $Keras$, and sample Python code.

\noindent\textbf{Text S1: Additional Relevance Propagation Rules}

We list additional relevance propagation rules here that are most relevant for LRP as we present it within the main manuscript, although we do not intent for this list to be comprehensive. For additional information, readers should refer to Chapter 10 of Samek et al. (2019), which provides a detailed discussion of the theory behind the various propagation rules developed thus far for LRP.

We mention within the main manuscript that we use a set of relevance propagation rules that only propagates information that increases the value of the selected output, although there are rules that permit the inclusion of information that reduces the value of the selected output as well. There are caveats to these additional rules, however, as the relevance is not conserved from the output value back to the input layer, and so interpretation of the relevance heatmaps becomes more subjective. We share the rules below for completeness, but we encourage the reader to be careful in their application and to carefully read the theory behind their development as provided by Bach et al. (2015), Montavon et al. (2017), and Samek et al. (2019)\nocite{bach2015pixel,montavon2017explaining,samek2019explainable}.

The relevance propagation rule that includes information that reduces the target output node is as follows:
\begin{equation}
    R_i = \sum_j \left(\alpha \frac{a_i w_{ij}^+}{\sum_i a_i w_{ij}^+} R_j - \beta \frac{a_i w_{ij}^-}{\sum_i a_i w_{ij}^-} \right)R_j.
\label{eqn:eqn1}
\end{equation}
Within Equation \ref{eqn:eqn1}, $\alpha$ represents the relative amount of positive relevance to be propagated backwards, $beta$ represents the relative amount of negative relevance to be propagated backwards, the $i$ subscript represents the $i$-th node in the layer of the network to which the relevance is being translated backwards, the $j$ subscript represents the $j$-th node in the layer of the network from which the relevance is being translated, $R_i$ is the relevance translated backwards to the $i$-th node, $R_j$ is the relevance of the $j$-th node, $a_i$ is the output from the $i$-th node after the non-linearity has been applied when the sample is passed forward through the network, $w_{ij}$ is the weight of the connection between the $i$-th and $j$-th nodes, and the $+$ and $-$ superscripts signifies that only positive or negative weights are considered, respectively.

There is also a separate propagation rule for relevance propagated backwards to the input layer from the first hidden layer for cases where the input is bounded between two values. In these cases, the relevance propagation rule is as follows:
\begin{equation}
R_i = \sum_j \left( \frac{x_iw_{ij} - lw_{ij}^+ - hw_{ij}^-}{\sum_i x_iw_{ij} - lw_{ij}^+ - hw_{ij}^-} \right)R_j.
\label{eqn:eqn2}
\end{equation}
Within Equation \ref{eqn:eqn2}, the $i$ subscript represents the $i$-th node in the layer of the network to which the relevance is being translated backwards, the $j$ subscript represents the $j$-th node in the layer of the network from which the relevance is being translated, $x$ represents the input value associated with the $i$-th node, $l$ represents the lower bound of the input dataset, $h$ represents the upper bound of the input dataset, $R_i$ is the relevance translated backwards to the $i$-th node, and $R_j$ is the relevance of the $j$-th node. This rule also conserves the total relevance translated from the first hidden layer backwards to the input layer, and abides by the rules of deep Taylor decomposition. If this rule is used in tandem with the rules presented within the manuscript that only propagate information backwards that increases the value of the output, then the relevance is conserved as it is propagated from the output node to the input layer. Additional information about deep Taylor decomposition is6 available within Montavon et al. (2017).

\noindent\textbf{Text S2: Additional Resources for LRP}

We offer a list of resources for programming LRP to be compatible with various $Python$ packages, although we do not intend for this list to be comprehensive. We use $innvestigate$, the first item in the below list, which is an implementation of LRP and various other neural network interpretation techniques for the $Keras$ neural network package within Python.

\begin{itemize}
    \item Python, $Keras/Tensorflow$ package: https://github.com/albermax/innvestigate \\
    \item Python, $Keras/Tensorflow$ package: https://github.com/nielsrolf/tensorflow-lrp \\
    \item Python, $Keras/Tensorflow$ package (coming soon): https://github.com/sicara/tf-explain \\
    \item Python, $Caffe$ package: https://github.com/sebastian-lapuschkin/lrp\_toolbox \\
    \item Python, $PyTorch$ package: https://github.com/moboehle/Pytorch-LRP \\
    \item MATLAB: http://www.heatmapping.org/lrptoolbox.html
\end{itemize}

%Delete all unused file types below. Copy/paste for multiples of each file type as needed.
\noindent\textbf{Code S1: Example Script for Optimal Input}

We provide an example script for the optimal input method using the $Keras$ package within Python. This is only meant to be an example, and the reader should be careful to understand each line of code within the script before attempting to implement it themselves. We only offer the code for the $Keras$ Python neural network package, although the concepts are transferable to any other neural network package, as well.

Example scripts using the $innvestigate$ package for LRP are available from the authors of the method at the following URL: $https://github.com/albermax/innvestigate$.

\clearpage

\renewcommand{\thefigure}{S1}

\begin{figure}[hp]
    \centerline{\includegraphics[width=30pc,angle=0]{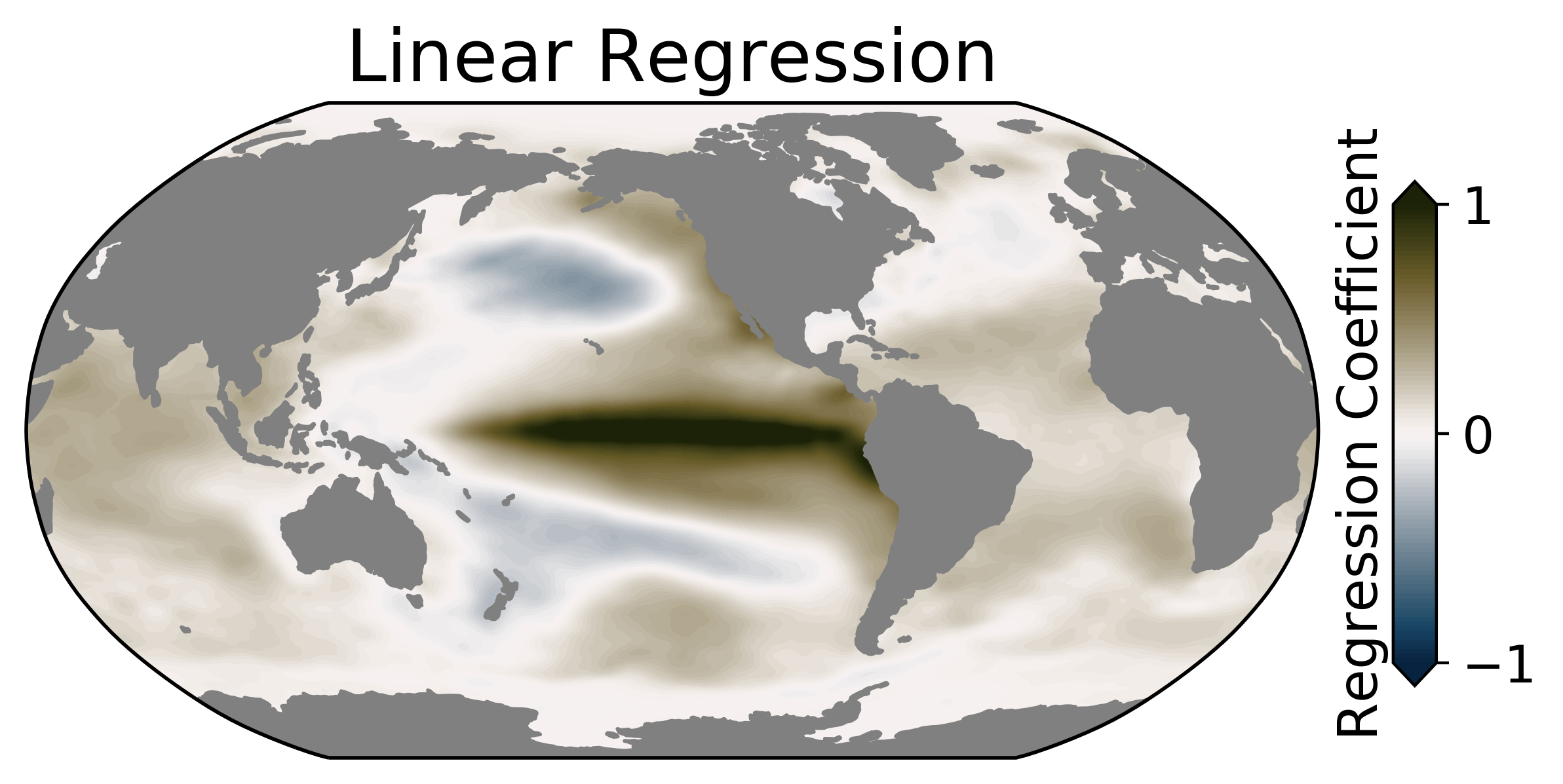}}
    \caption{Regression coefficients for the linear regression method to predict the phase of ENSO using global maps of sea-surface temperature anomalies. The regression coefficients are calculated by regressing the time series of global sea-surface temperature anomaly maps onto the standardized ENSO time series. We then project this map of regression coefficients onto the global sea-surface temperature anomalies to predict the sign of ENSO phase. The resultant accuracy predicting the ENSO phase using the regression coefficients is 81.5\%.}
    \label{fig:fig6}
\end{figure}

\clearpage

\renewcommand{\thefigure}{S2}

\begin{figure}[hp]
    \centerline{\includegraphics[width=22pc,angle=0]{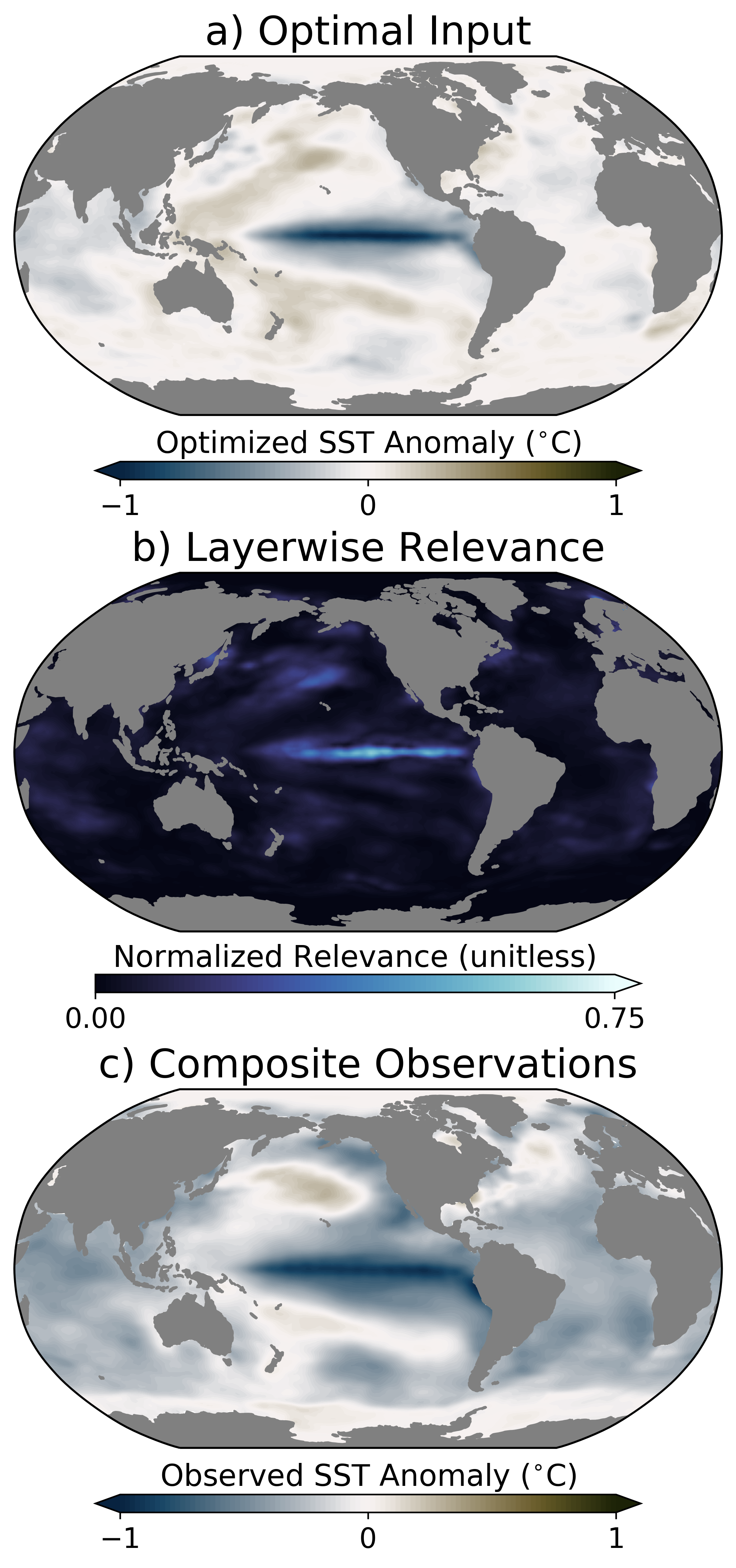}}
    \caption{As in Figure 6 of the main manuscript, but for the neural network's understanding of the spatial structure of La Ni\~na based on a total of 485 samples (including both testing and training data).}
    \label{fig:fig6}
\end{figure}

\clearpage

\renewcommand{\thefigure}{S3}

\begin{figure}[hp]
    \centerline{\includegraphics[width=26pc,angle=0]{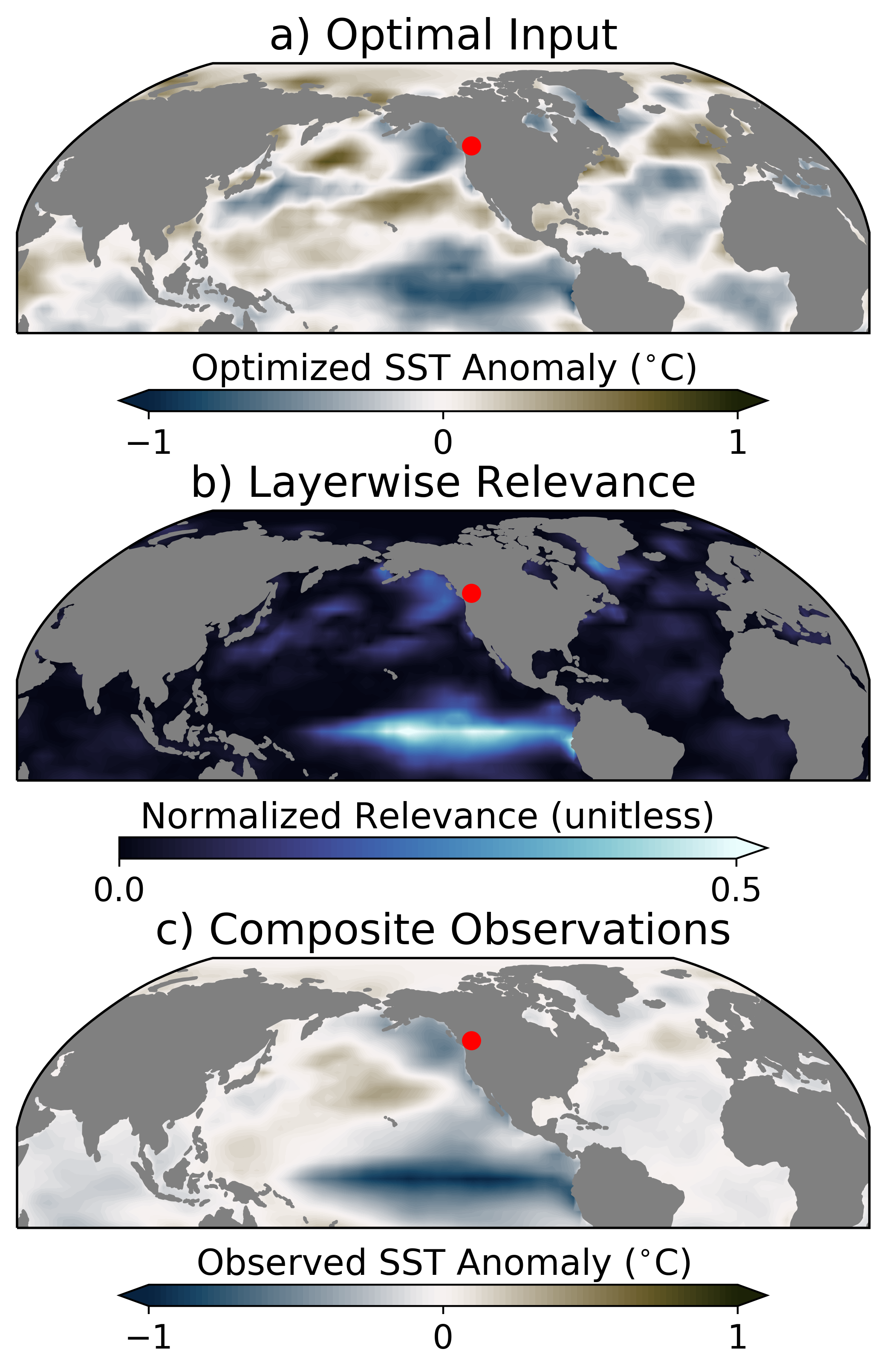}}
    \caption{As in Figure 9 of the main manuscript, but for the neural network's understanding of the sea-surface temperature anomalies that lend predictability for \textit{negative} surface temperature anomalies at the red dot.}
    \label{fig:fig9}
\end{figure}

\end{article}
\clearpage